\documentclass[prl,superscriptaddress,twocolumn,eqsecnum,byrevtex]{revtex4}
\usepackage{amsfonts}
\usepackage{amssymb}
\usepackage{amsmath}
\usepackage{color}
\usepackage{graphicx}
\usepackage{subcaption}
\usepackage{time}
\usepackage[colorinlistoftodos]{todonotes}
\usepackage{sidecap}
\usepackage{chngcntr}
\begin{document}
%
%
\title{Stochastic Variational Approach to Small Atoms and Molecules Coupled to Quantum Field Modes}

\author{Alexander Ahrens}
\affiliation{Department of Physics and Astronomy, Vanderbilt University, Nashville, Tennessee, 37235, USA}
\author{Chenhang Huang}
\affiliation{Department of Physics and Astronomy, Vanderbilt University, Nashville, Tennessee, 37235, USA}
\author{Matt Beutel}
\affiliation{Department of Physics and Astronomy, Vanderbilt University, Nashville, Tennessee, 37235, USA}
\author{Cody Covington}
\affiliation{Department of Chemistry, Austin Peay State University, Clarksville, USA}

\author{K\'alm\'an Varga}
\email{kalman.varga@vanderbilt.edu}
\affiliation{Department of Physics and Astronomy, Vanderbilt University, Nashville, Tennessee, 37235, USA}

\begin{abstract}
In this work, we present a stochastic variational calculation (SVM) of energies and wave functions of few particle systems coupled to quantum fields in cavity QED. The light-matter coupled system is described by the Pauli–Fierz Hamiltonian. The spatial wave function and the photon spaces are optimized by a random selection process. Examples for a two-dimensional 
trion and confined electrons as well as for the He atom and
the H$_2$ molecule are presented showing that the light-matter coupling  drastically changes the electronic states. 
\end{abstract}

\maketitle
\counterwithout{equation}{section}
Strong coupling of cavity electromagnetic modes and molecules create hybrid 
light-matter states that combine the properties of their ingredients 
modifying potential energy surfaces, charge states, and electronic structure.  
The possibility of altering physical and chemical properties by coupling to light attracted 
intense experimental \cite{PhysRevLett.114.196402, Balili1007,PhysRevLett.114.196403,Xiang665,
doi:10.1021/acsphotonics.0c01224,Coles2014,Kasprzak2006,PhysRevLett.106.196405,Plumhof2014,
https://doi.org/10.1002/adma.201203682,Wang2021,BasovAsenjoGarciaSchuckZhuRubio+2021+549+577} 
and theoretical interest\cite{PhysRevLett.122.017401,DiStefano2019,PhysRevX.5.041022,
PhysRevLett.116.238301,Galego2016,Shalabney2015,doi:10.1021/acsphotonics.9b00648,Schafer4883,
Ruggenthaler2018,Flick15285,Flick3026,PhysRevLett.123.083201,Mandal,doi:10.1021/acsphotonics.9b00768,
doi:10.1021/acs.nanolett.9b00183,FlickRiveraNarang,PhysRevLett.121.113002,Garcia-Vidaleabd0336,
Thomas615,PhysRevResearch.2.023262,doi:10.1063/5.0036283,doi:10.1063/5.0038748,doi:10.1063/5.0039256,
doi:10.1063/5.0012723,doi:10.1063/5.0021033,acs.jpcb.0c03227,PhysRevLett.119.136001,
doi:10.1063/5.0012723,Flick15285,doi:10.1021/acsphotonics.7b01279,PhysRevA.98.043801,
doi:10.1021/acs.jpclett.0c01556,doi:10.1021/acs.jctc.0c00618,doi:10.1021/acs.jpclett.0c03436,
doi:10.1021/acs.jctc.0c00469,PhysRevB.98.235123,PhysRevLett.122.193603,
Szidarovszky_2020,doi:10.1021/acs.jpclett.1c01570}.
The experimental investigations are overarching exciton transport \cite{PhysRevLett.114.196402,PhysRevLett.114.196403}, 
polariton condensation \cite{Balili1007,Plumhof2014}, transfer of excitation 
\cite{Coles2014}, and chemical reactivity \cite{https://doi.org/10.1002/anie.201605504}.
The theoretical works  explored excitation and charge transfer
\cite{Schafer4883}, self-polarization 
\cite{doi:10.1063/5.0012723}, potential energy surfaces \cite{PhysRevLett.123.083201} electron transfer 
\cite{acs.jpcb.0c03227}, excitons \cite{doi:10.1021/acs.nanolett.9b00183}, 
ionization potentials \cite{doi:10.1063/5.0038748}, and intermolecular interactions 
\cite{doi:10.1063/5.0039256} in cavities, to name a few. 

For atoms and molecules in a vacuum, high precision measurements and theoretical calculations have been developed.
For example, the accuracy of theoretical prediction 
\cite{PhysRevLett.122.103003} and experimental measurement
\cite{PhysRevLett.122.103002} reaches the level of 1 MHz for the dissociation energy of the H$_2$ molecule.
The theoretical description of the light-matter hybrid 
states is far from this accuracy. The main reason behind this is that the accurate wave 
function methods developed in quantum chemistry are tedious and computationally expensive 
even before an additional degree of freedom 
(light) is added, and the density functional theory 
\cite{PhysRev.140.A1133} calculations lack suitable exchange 
correlation functionals for light-matter coupling. Initial developments have been started 
in both directions \cite{doi:10.1021/acsphotonics.9b00648,Flick15285,doi:10.1021/acsphotonics.9b00768,PhysRevResearch.2.023262,
doi:10.1063/5.0038748,doi:10.1063/5.0021033}.

In this work, we use the stochastic variational method to build
optimized light-matter coupled wave functions. The calculations
can reach the same accuracy as conventional high precision calculations
for small systems. We will show that the light-matter coupled wave
functions are  drastically different from the non-coupled
electronic wave functions and the wave functions belonging to different
photon spaces are very dissimilar (requiring separate optimizations). 

The spatial wave functions will be represented by
Explicitly Correlated  Gaussian (ECG) basis functions \cite{RevModPhys.85.693}. 
The advantage of  the approach is that the matrix elements are analytically 
available
\cite{suzuki1998stochastic,doi:10.1063/1.4974273,Zaklama2019} and  it
allows very accurate calculations of energies and wave functions
\cite{RevModPhys.85.693,Zhang2015,PhysRevB.93.125423,PhysRevB.61.13873,
PhysRevLett.83.5471,PhysRevB.59.5652,PhysRevB.63.205308}.

We assume that the system is nonrelativistic and the coupling to the light can be described by the dipole approximation.  
The Pauli-Fierz (PF) non-relativistic QED Hamiltonian 
provides a consistent quantum description at this level 
\cite{Ruggenthaler2018,Rokaj_2018,Mandal,acs.jpcb.0c03227,PhysRevB.98.235123}. 
The PF Hamiltonian in the Coulomb gauge is
$H=H_e+H_{ep}$
where $H_e$ is the electronic Hamiltonian and
\begin{equation}
H_{ep}=\sum_{\alpha=1}^{N_p}\left[ \omega_{\alpha}\left(\hat{a}_{\alpha}^{+} \hat{a}_{\alpha}+\frac{1}{2}\right)
-\omega_{\alpha}q_\alpha
    \vec{\lambda}_{\alpha}\vec{D}+{\frac{1}{2}}  \left(\vec{\lambda_{\alpha}}
\vec{D}\right)^{2}\right],
\label{hep}
\end{equation}
(atomic units are used in this work). In Eq. \eqref{hep}  $\vec{D}$ is the dipole operator, 
the photon fields are described by quantized oscillators, and
$q_\alpha={\frac{1}{ \sqrt{2\omega_\alpha}}}(\hat{a}^+_\alpha+\hat{a}_\alpha)$ 
is the displacement field.
This  Hamiltonian describes $N_p$ photon modes with photon frequency
$\omega_{\alpha}$ and coupling  $\vec{\lambda}_{\alpha}$.  
The coupling term is usually written as \cite{PhysRevA.90.012508} 
$\vec{\lambda}_{\alpha}=\sqrt{4\pi}\,S_\alpha(\vec{r})\vec{e}_\alpha$,
where $S_\alpha(\vec{r})$ is the cavity mode function at position $\vec{r}$
and $\vec{e}_\alpha$  is the transversal polarization vector of the photon modes.
The first term in Eq. \eqref{hep} is the Hamiltonian of the photon modes, the second term couples the photons to the dipole, and the last term is the dipole self-interaction,
$H_d={1\over 2} \sum_{\alpha=1}^{N_p}\left(\vec{\lambda_{\alpha}} \vec{D}\right)^{2}$.

The Hamiltonian of an $N$ electron system interacting with a Coulomb interaction and confined in 
an external potential $V_c$ reads as
\begin{equation}
    H_e=-\sum_{i=1}^N {{\nabla^2_i}\over 2 m_i}+\sum_{i<j}^N
    {q_iq_j\over \vert \mathbf{r}_i-\mathbf{r}_j\vert}+\sum_{i=1}^N V_c(\mathbf{r}_i),
\end{equation}
where $\mathbf{r}_i,q_i$, and $m_i$ are the coordinate, charge, and mass of the $i$th particle ($m_i=1$ for electrons in atomic units).

Introducing the shorthand notations  $\vec{r}=(\mathbf{r}_1,...,\mathbf{r}_N)$, and
$
\vert\vec{n}\rangle = \vert n_1\rangle\vert n_2\rangle,{\ldots},\vert
n_{N_p}\rangle
$
where $n_\alpha$ is  the number of photons in photon mode
$\alpha$, the variational trial wave function is written as a linear
combination of  products of spatial and photon
space basis functions
\begin{equation}
\Psi(\vec{r})=\sum_{\vec{n}}\sum_{k=1}^{K_{\vec{n}}}
    c_k^{\vec{n}}\psi_k^{\vec{n}}(\vec{r})|\vec{n}\rangle.
    \label{pwfM}
\end{equation}

The spatial part of the wave function is expanded into ECGs 
for each photon state $|\vec{n}\rangle$ as
\begin{equation}
    \psi_k^{\vec{n}}(\vec{r})={\cal A}\lbrace {\rm e}^{-{1\over 2}\sum_{i<j}^N
    \alpha_{ij}^k(\mathbf{r}_i-\mathbf{r}_j)^2-{1\over 2}\sum_{i=1}^N \beta_i^k(\mathbf{\mathbf{r}}_i-\mathbf{s}_i^k)^2}
    \Lambda({\vec{r}})
    \chi_S\rbrace
\label{bf}
\end{equation}
where ${\cal A}$ is an antisymmetrizer, $\chi_S$ is the $N$ electron spin function (coupling the spin to $S$), and 
$\alpha_{ij}^k$,$\beta_i^k$ and $\mathbf{s}_i^k$ are nonlinear parameters.

The dipole self-energy introduces a non-spherical term 
into the Hamiltonian. The solution of this nonspherical problem is very
difficult and slowly converging using angular momentum expansions. 
To avoid this we introduce  $\Lambda(\vec{r})={\rm e}^{\vec{r}U\vec{r}}$
in  Eq. \eqref{bf} to eliminate the dipole self-interaction term $H_d$
from Eq. \eqref{hep} altogether. In the exponential, $U$ is a $3N\times 3N$ matrix
with elements chosen in such a way that when the kinetic energy acts
on the trial function, the resulting expression cancels the dipole self-energy \cite{Supp}. 

The necessary matrix elements can be analytically calculated for both
the spatial \cite{suzuki1998stochastic,RevModPhys.85.693}  and the
photon parts and the resulting Hamiltonian and overlap matrices are very
sparse matrices \cite{Supp}.

We will optimize the basis functions selecting the best spatial
basis parameters and photon components using the SVM.
In the SVM, the basis functions are optimized by randomly generating  a large 
number of candidates and selecting the ones that 
give the lowest energy \cite{RevModPhys.85.693,suzuki1998stochastic,kukulin77}. The size 
of the basis can be increased by adding the best states one by one 
and a $K$ dimensional basis can be refined by replacing states with randomly 
selected better basis functions. This approach is very efficient in 
finding suitable basis functions. The details of the SVM selection are
given in \cite{Supp}. 

\begin{figure}
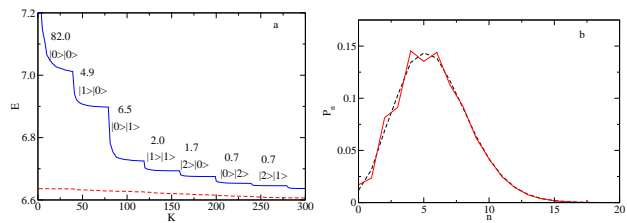

\begin{minipage}{0.5\textwidth}
\includegraphics[width=0.45\textwidth]{figure1a.eps}
\includegraphics[width=0.45\textwidth]{figure1b.eps}
\end{minipage}
\caption{(a) Energy convergence of  the SVM calculation,
(b)Photon space occupations for  the ground state with
$\vec{\lambda}=(0.5,0.5)$ a.u., solid line: SVM, dashed line:
exact diagonalization.}
\label{conv}
\end{figure}

In Fig. \ref{conv}a we show an example for the energy convergence of a two
dimensional trion (2 electrons and a hole) coupled to 2 photon modes using the
SVM with $K_{\vec n}$=40 basis vectors in each photon state. Initially the
photon space $|n_1\rangle|n_2\rangle$ is restricted by $n_1+n_2<4$. 
First the $|0\rangle|0\rangle$ space is optimized by
adding spatial basis functions selected by SVM one by one. Then the
$|1\rangle|0\rangle$ space is added and populated by SVM and this
process is repeated for each allowed photon space.
The energy quickly converges in the photon spaces and the energy
gain is less and less as higher photon numbers are added.
The energy is not fully converged in the $|0\rangle|0\rangle$ space so this space
may need more basis functions, but in other spaces, the convergence is close
to perfect and smaller spatial basis can be used. This is remedied by
the refining step \cite{Supp}, where the basis is improved by randomly 
replacing photon spaces with energetically more favorable ones. In
this step, the photons can also couple photon spaces with $n_1+n_2\ge
4$. As the dashed line in Fig. \ref{conv}a. shows, this step still
significantly improves the energy. 

To test the accuracy of the approach we have compared the 
converged energies to a one-electron one-photon problem which 
can be solved with finite difference representation of the 
spatial wave function and exact diagonalization of the coupled system
\cite{Flick15285,doi:10.1021/acsphotonics.7b01279,PhysRevA.98.043801}.
The SVM and exact diagonalization energies agree up to 5 digits
for the lowest five states \cite{Supp}. Fig. \ref{conv}b shows photon 
occupation probabilities for a very strong coupling case. The SVM and the exact diagonalization 
results are in very good agreement, and the slight discrepancy might be
due to the finite differences used in the exact diagonalization
approach. 

We use a harmonically confined  ($V_c(r)={1\over 2}\omega_c^2 r^2$)
two-dimensional (2D) spin singlet two-electron system  
coupled to 2 photon modes as a second test case. 
The spatial part of this problem is analytically solvable and the coupled light-matter 
Hamiltonian can be
diagonalized \cite{henry}. Using $\omega_c=1$ a.u. for the harmonic
confinement, $\omega_1=1,\omega_2=1$ a.u. for the photon frequencies 
and $\vec{\lambda}_1=(1,1),\vec{\lambda}_2=(-1,-1)$ a.u. for the coupling,
the exact energy of the system is $E$=3.65400 a.u. This choice corresponds
to a diagonally polarized light and mode functions
$S_\alpha(\vec{r})=\sqrt{2\over A L}\sin(k_\alpha x)$ where $L$ is the length, $AL$ is the volume of the cavity, and $k_\alpha=\alpha \pi/L$ is the wave vector
($\alpha=1,2$). Our system is placed at the center of the cavity,
$x=L/2$ and ${\vec{\lambda}}_2=-{\vec{\lambda}}_1$. The calculated energy
$E=$3.65401 a.u. is in perfect agreement with the analytical value and
the photon space probabilities also agree \cite{Supp}. This is a
good and difficult test for the numerical calculation because 
two different couplings and many (over 80) photon states are present
and photon states with very small occupation numbers contribute to the
energy.
\begin{figure}
\begin{minipage}{0.5\textwidth}
\includegraphics[width=0.45\textwidth]{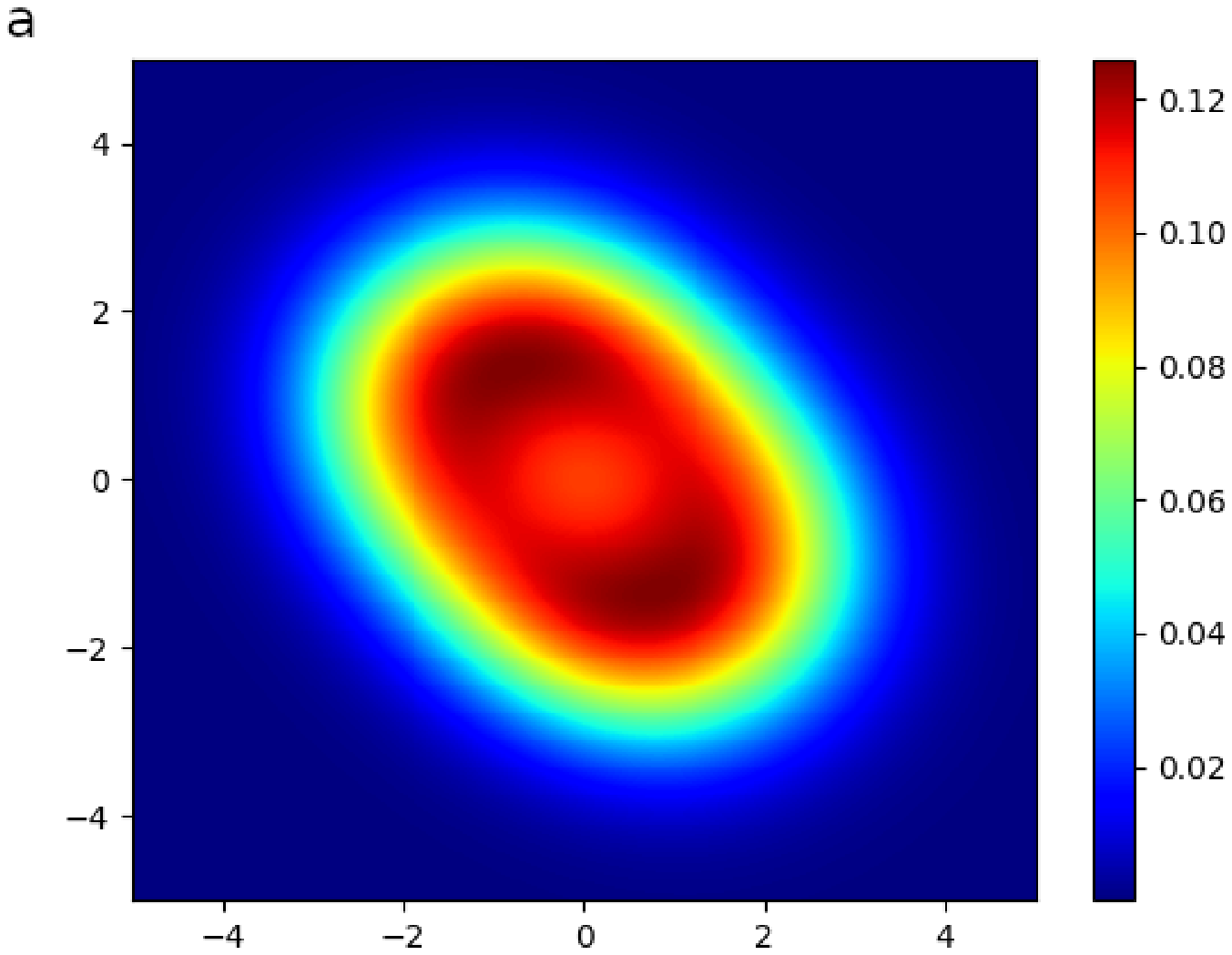}
\includegraphics[width=0.45\textwidth]{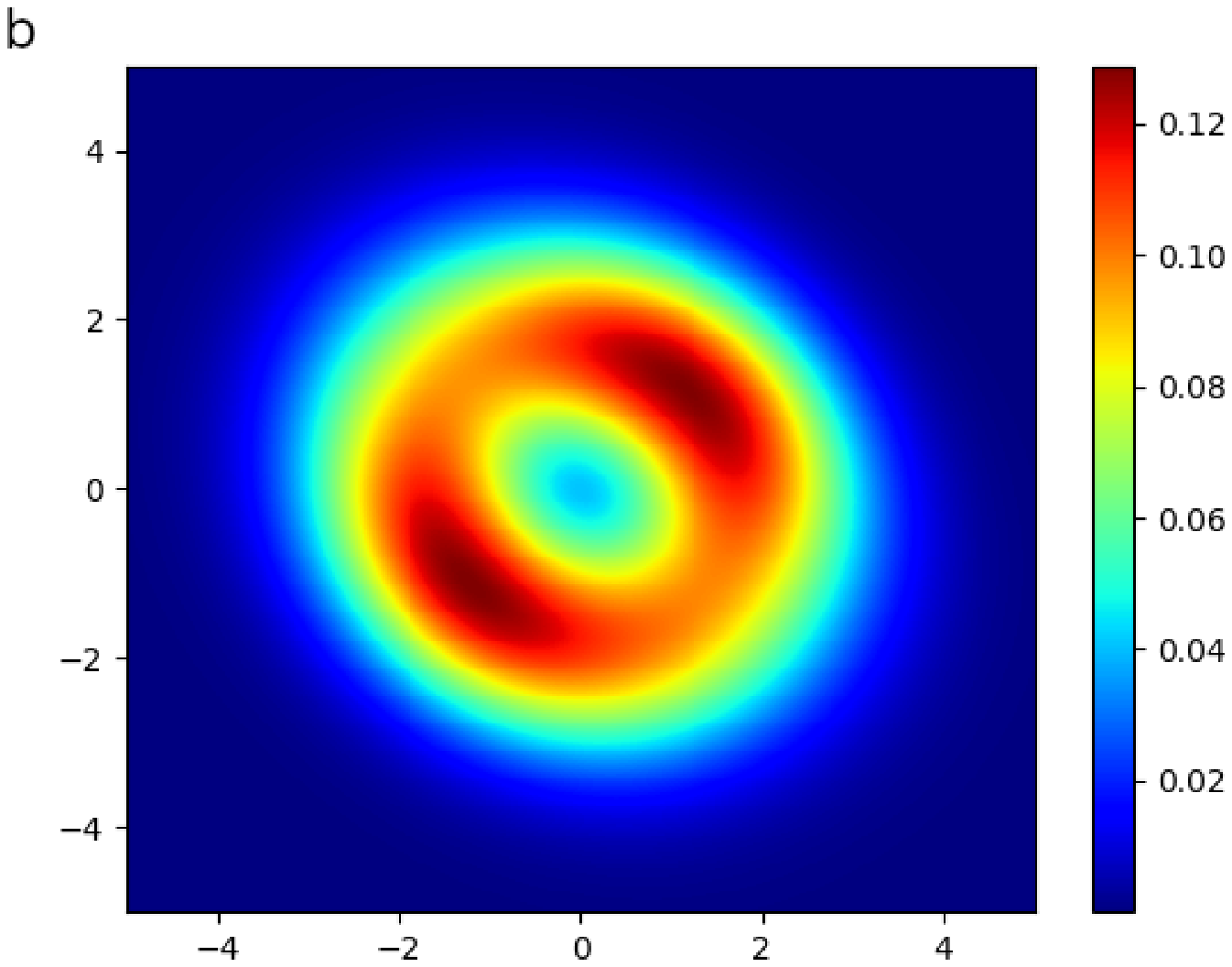}
\end{minipage}
\caption{Electron density of a harmonically confined 2D three-electron
system ($\omega=2$ and $\lambda=4$) (a) S=1/2, (b) S=3/2.}
\label{3e}
\end{figure}

Next we study a 2D 3 electron system in harmonic confinement 
($\omega_c=1/3$ a.u.) coupled to  photon mode of $\omega_1=2$ and
$\vec{\lambda}_1=(4,4)$.
We use two dimensional examples because the visualization of the density is simpler. Without
coupling to the light the electron density of these systems is spherically symmetric. The S=1/2 density has a peak in the center,
while the $S=3/2$ system forms a ring-like structure due to the Pauli
repulsion between the spin-polarized electrons \cite{Supp}.  
The electron density is shown in Fig. \ref{3e} for $S=1/2$ and $S=3/2$ with a very different structure 
for the coupled case. The density peak
in the S=1/2 case breaks up into two peaks and  the S=3/2 ring also 
splits into two parts. This shows how strongly the light-matter
coupling can change the electron density of the system.
\begin{figure}
\begin{minipage}{0.5\textwidth}
\includegraphics[width=0.3\textwidth]{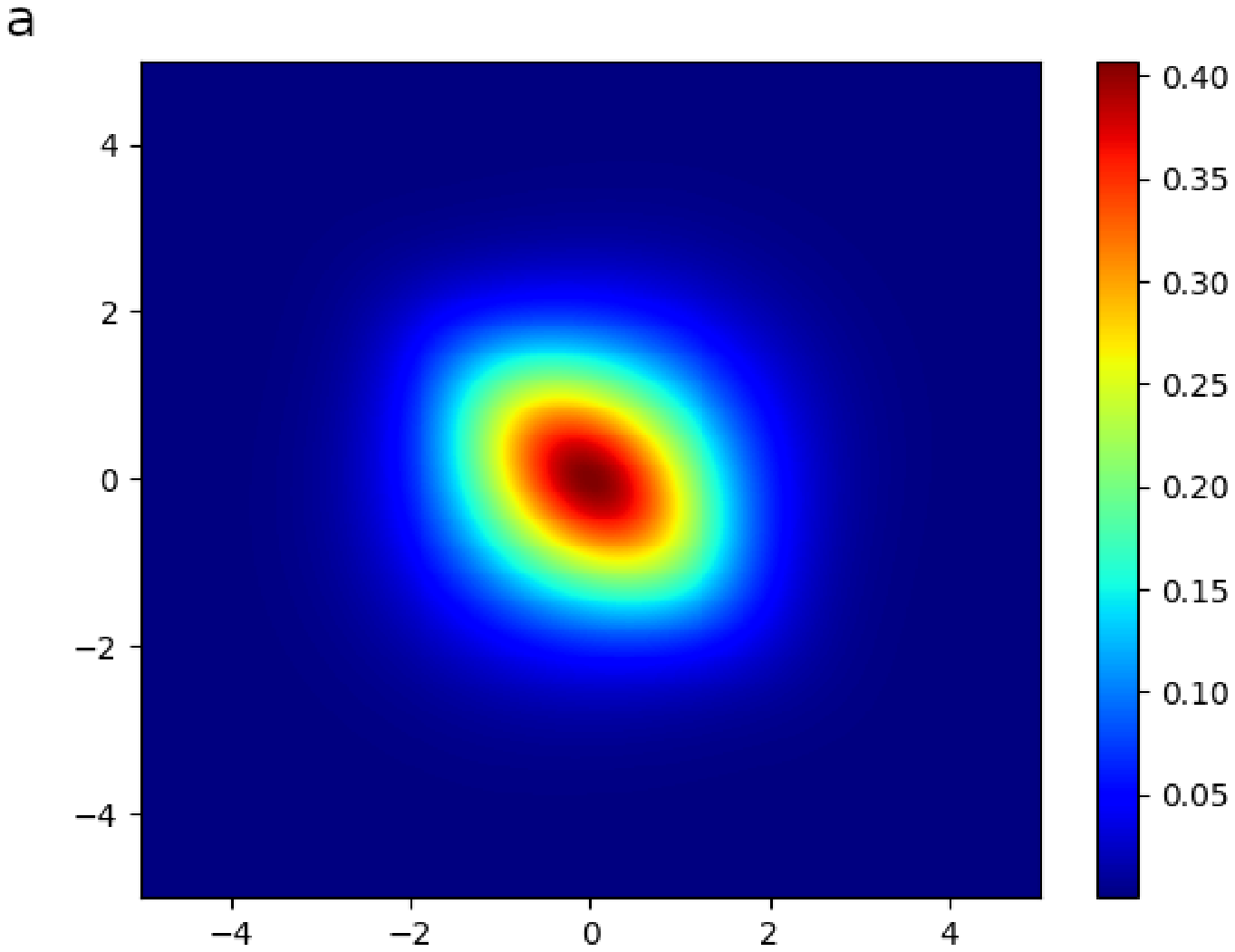}
\includegraphics[width=0.3\textwidth]{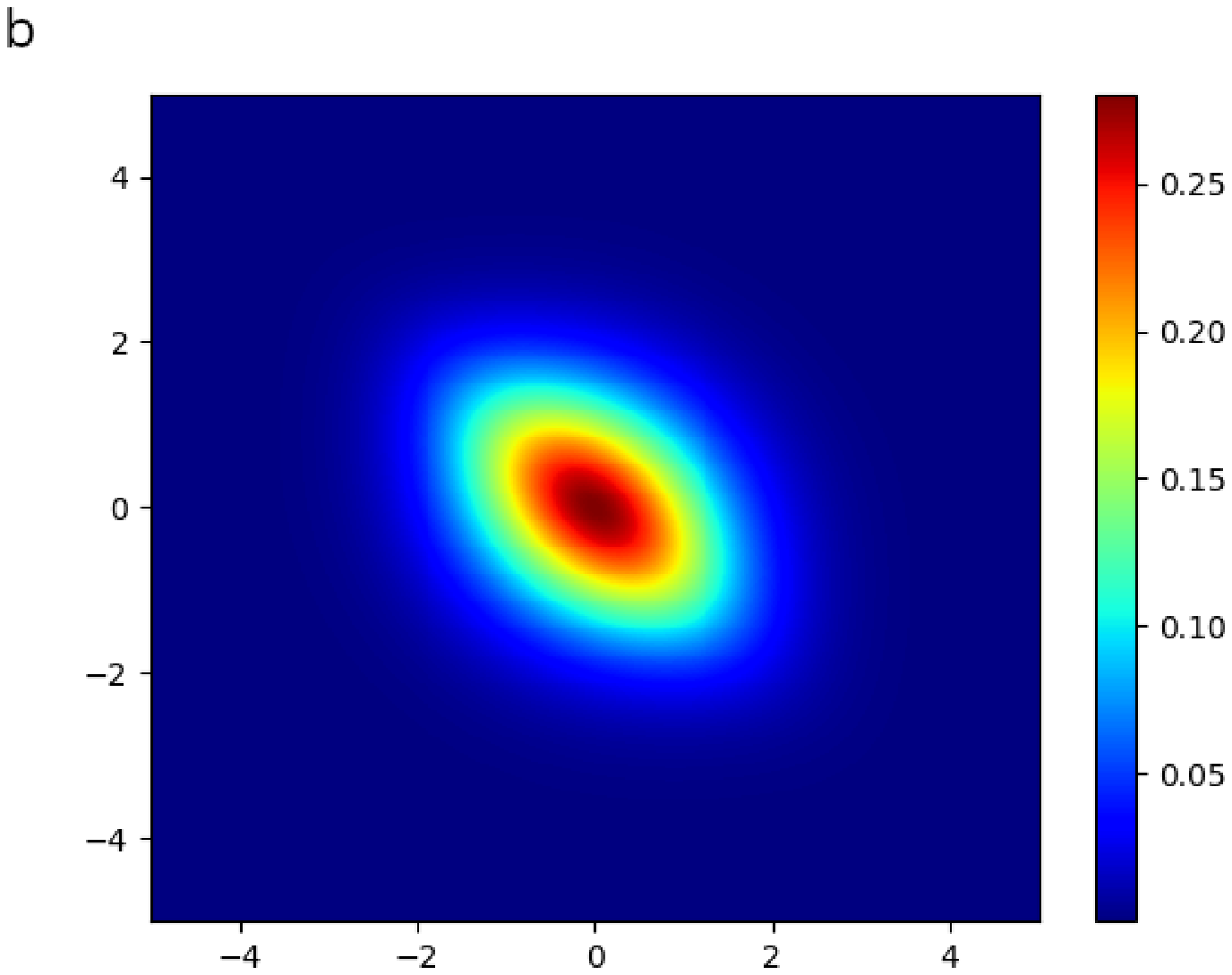}
\includegraphics[width=0.3\textwidth]{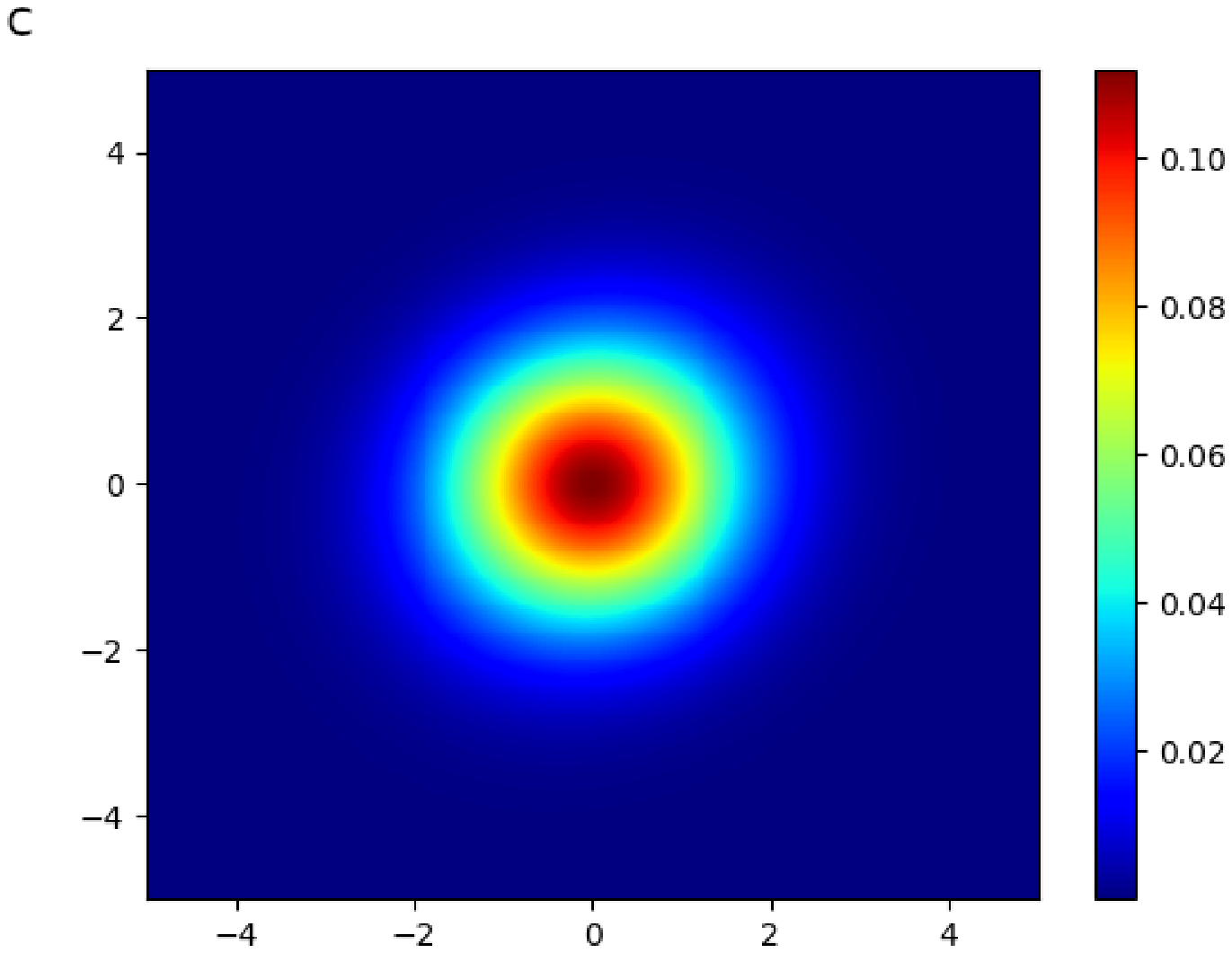}\\\quad
\includegraphics[width=0.3\textwidth]{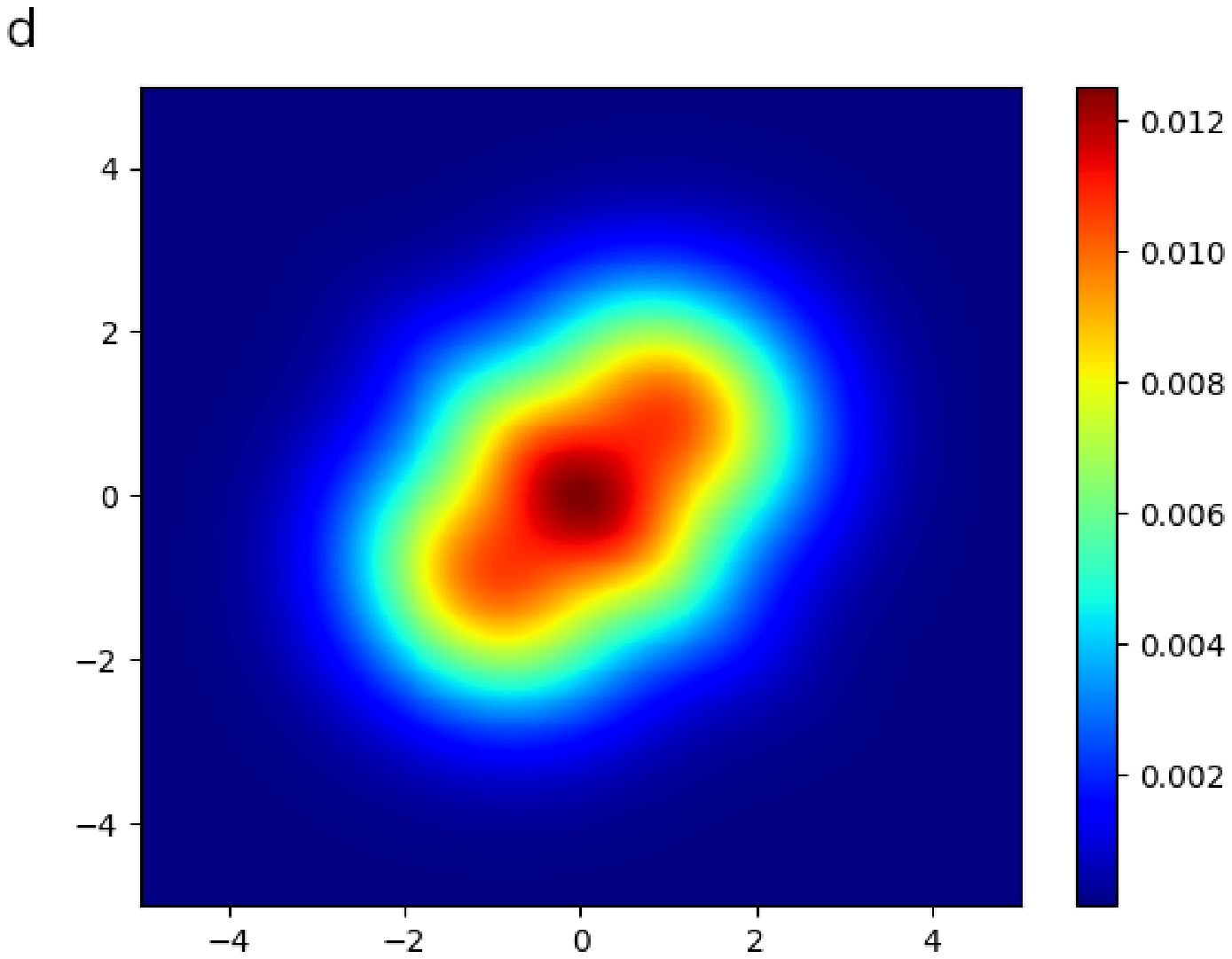}
\includegraphics[width=0.3\textwidth]{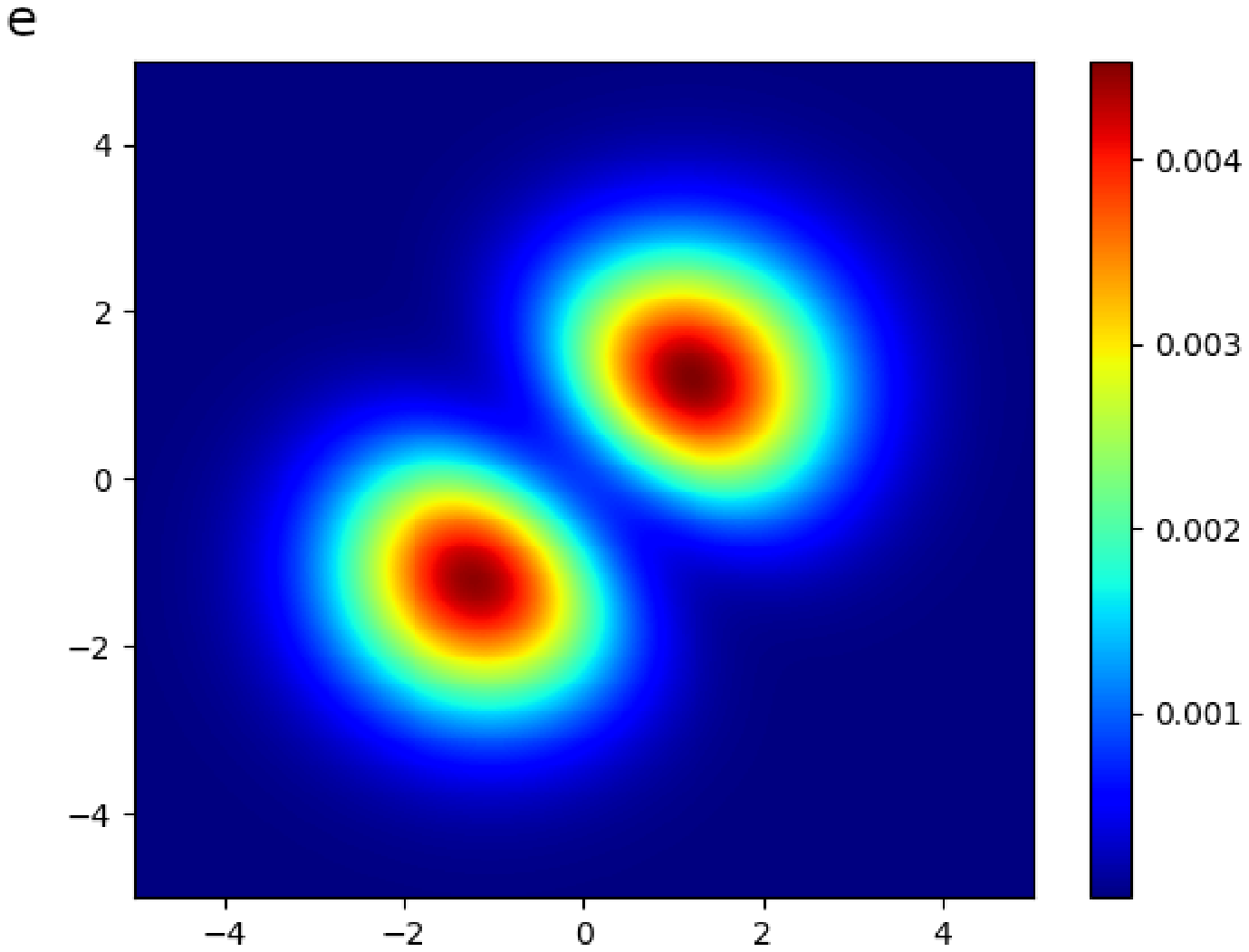}
\includegraphics[width=0.3\textwidth]{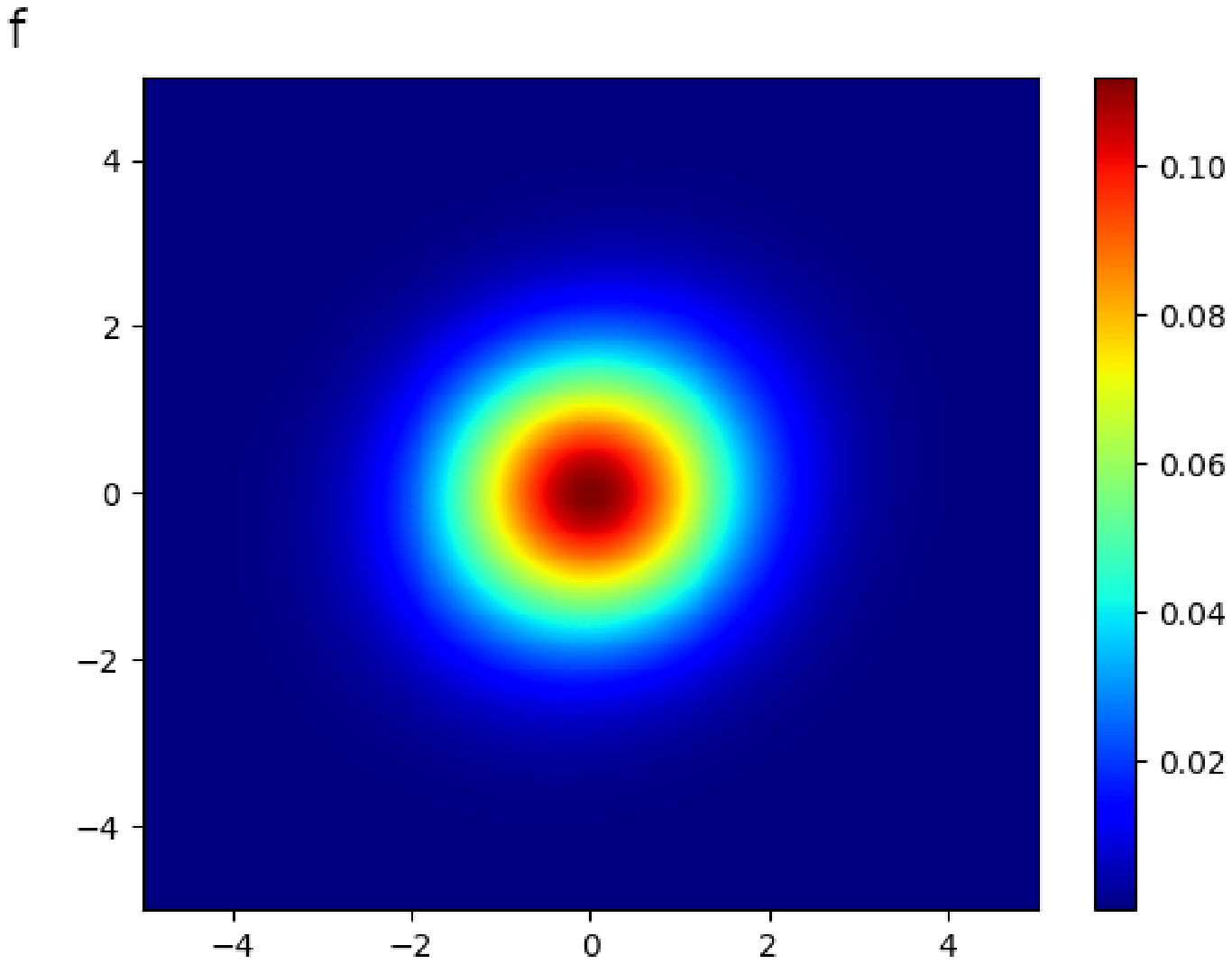}\\\quad
\includegraphics[width=0.3\textwidth]{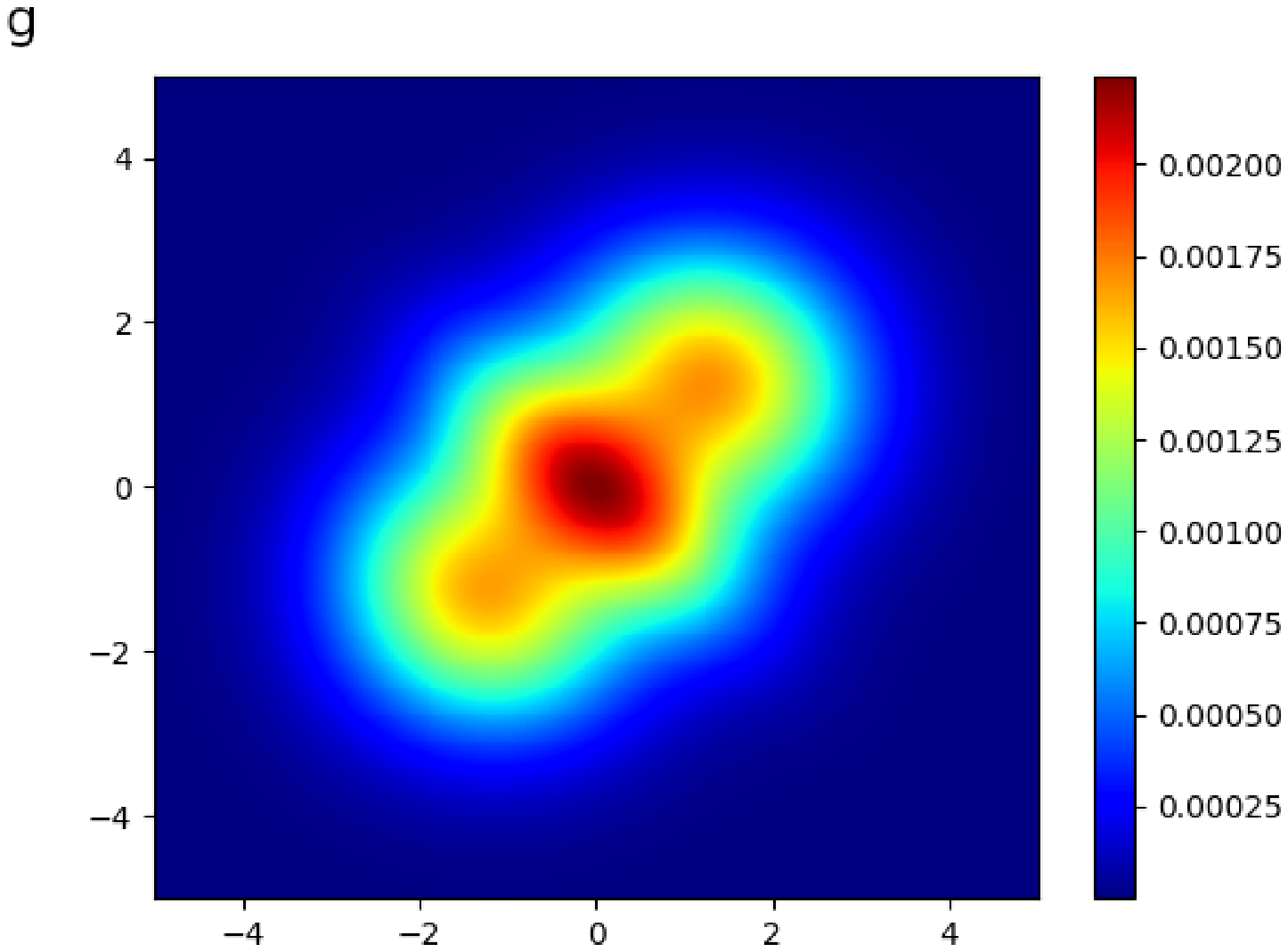}
\includegraphics[width=0.3\textwidth]{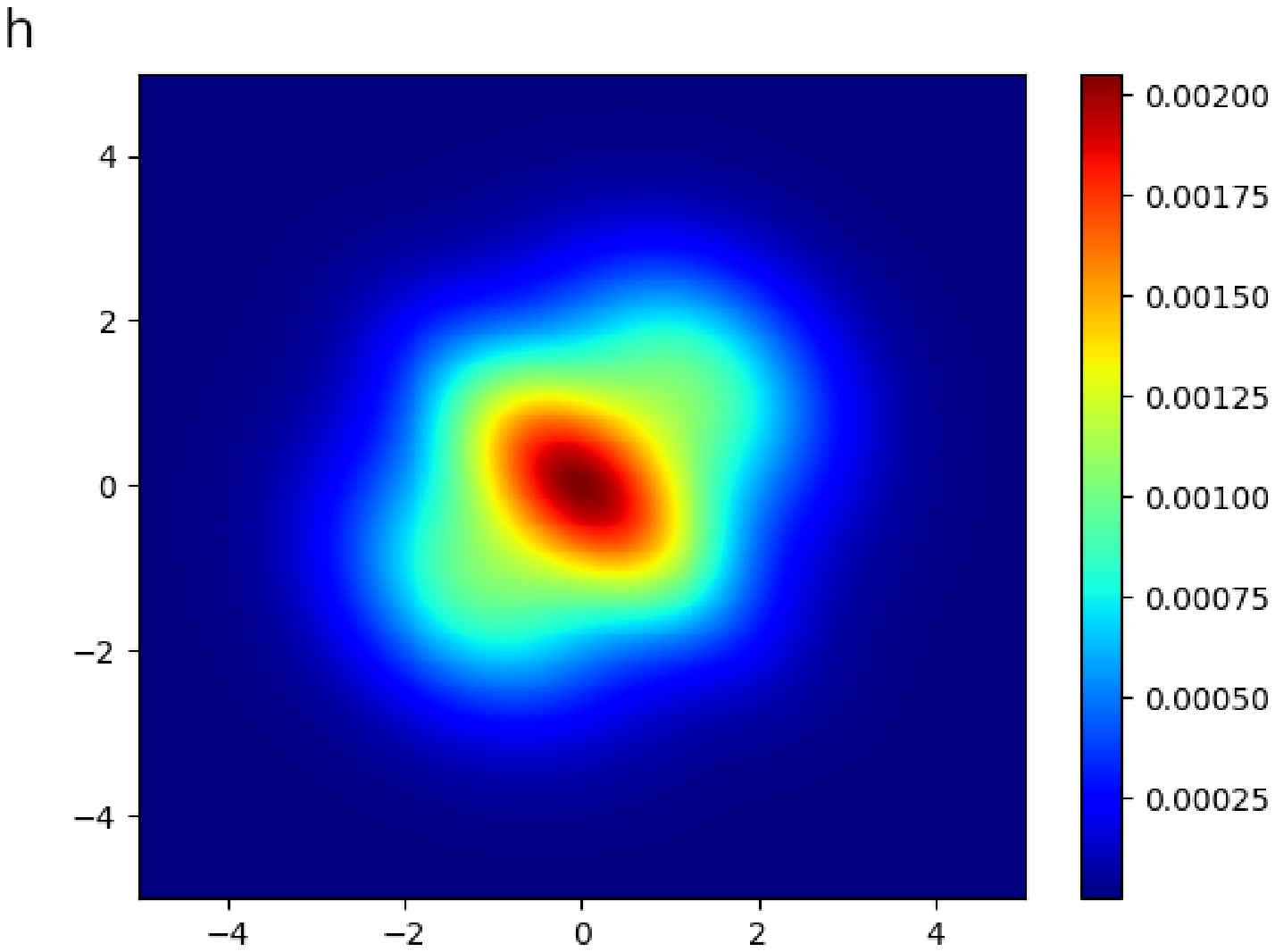}
\includegraphics[width=0.3\textwidth]{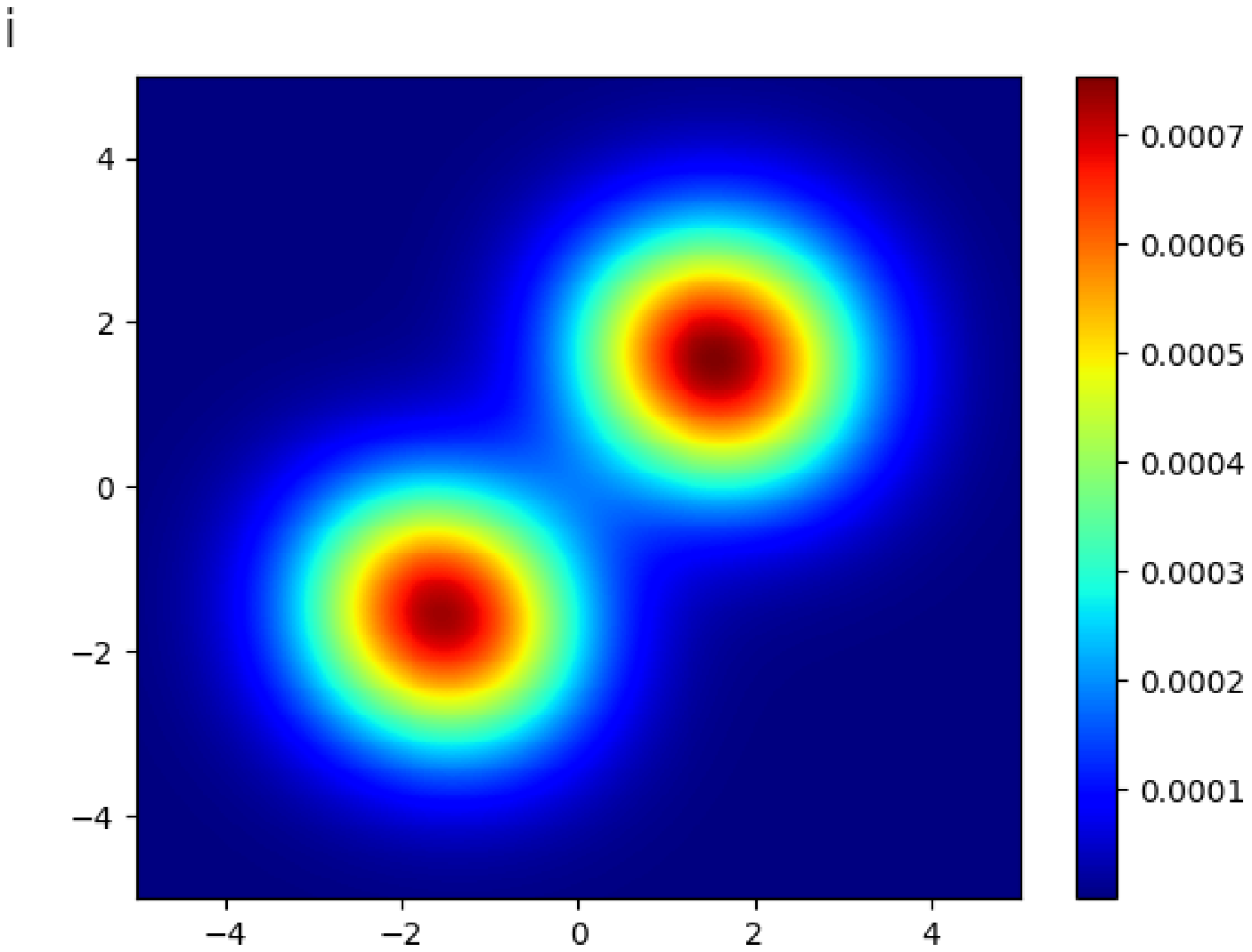}\\\quad
\includegraphics[width=0.3\textwidth]{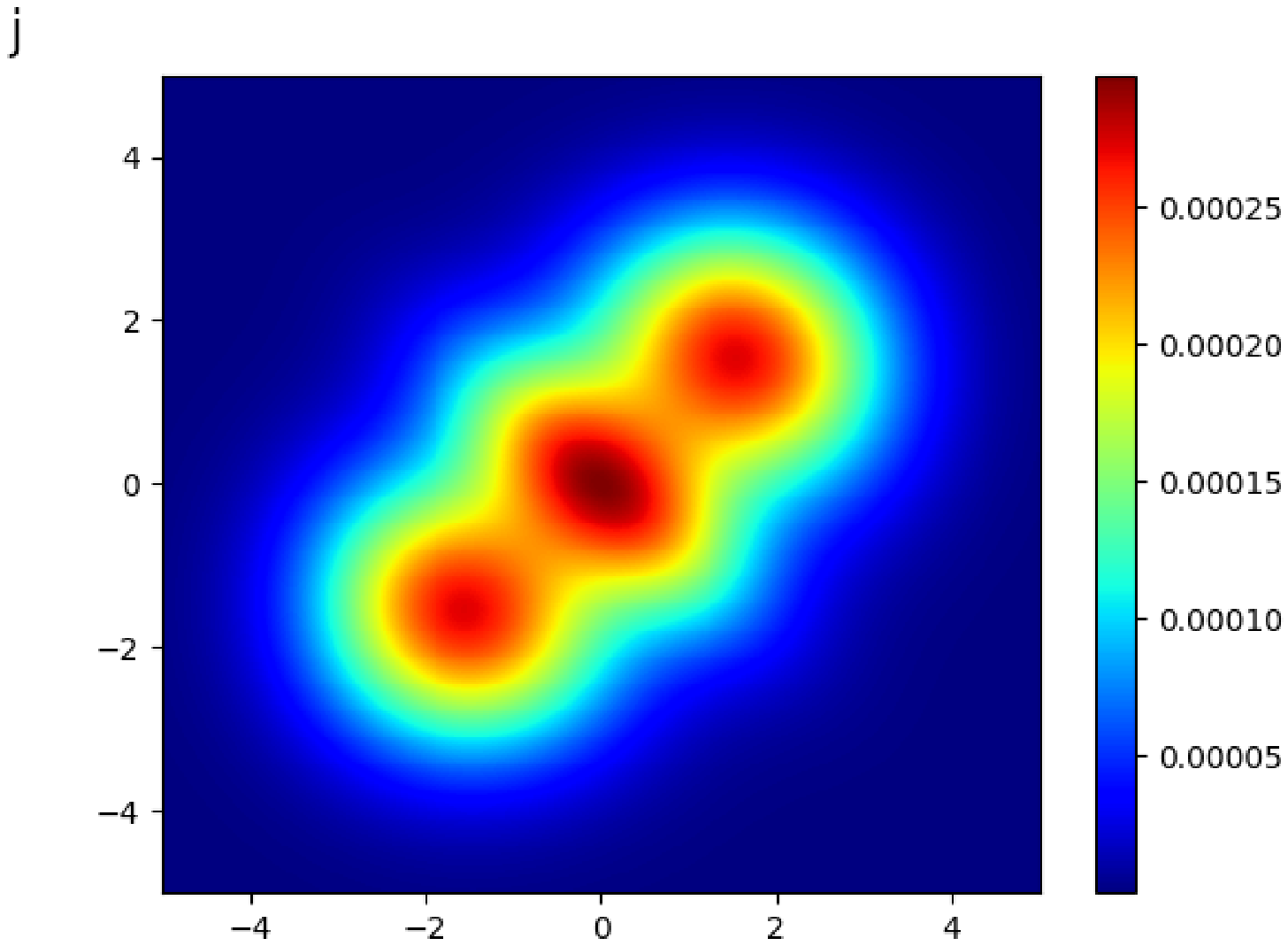}
\includegraphics[width=0.3\textwidth]{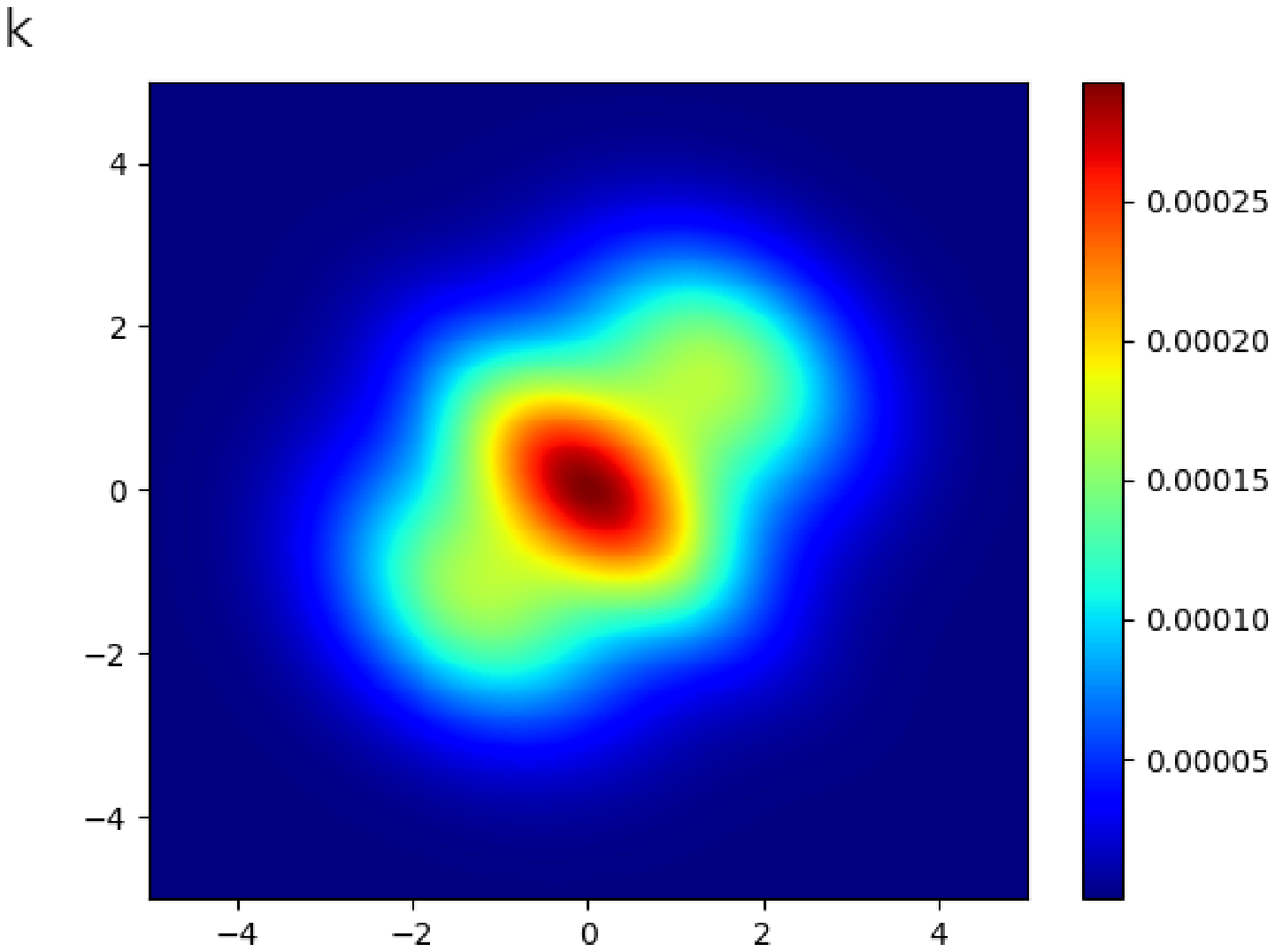}
\includegraphics[width=0.3\textwidth]{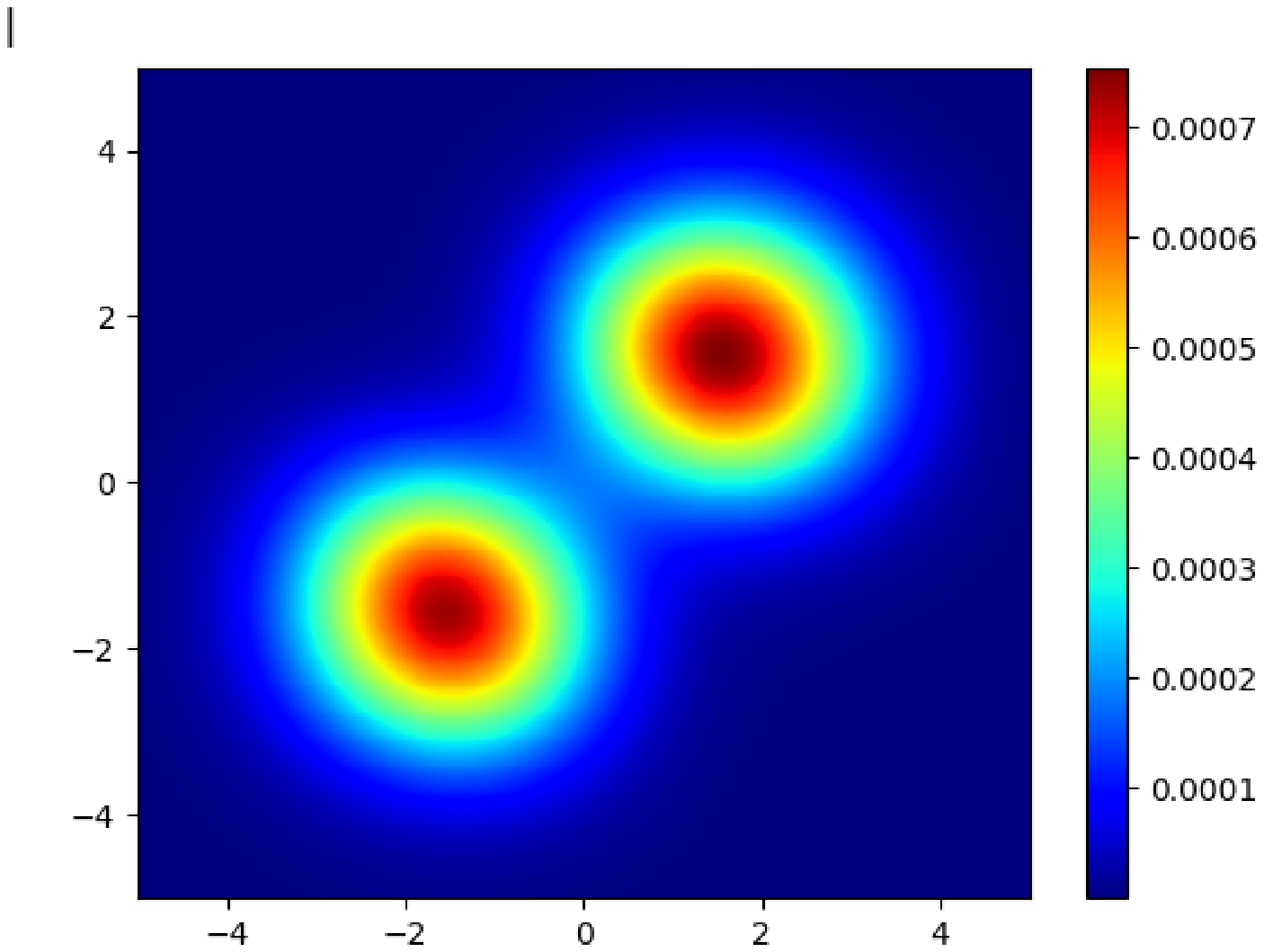}\\\quad
\end{minipage}
\caption{Two electrons and a hole in 2D coupled to 1 photon mode
($\omega_1=2,\vec{\lambda}_1=(4,4)$ a.u.). The left column shows the 
total single particle density; the electron density is in the middle; and the right column shows the density of the hole. The first row
belongs to the photon space $\vert 0\rangle$, the second to 
$\vert 1\rangle$, the third to  $\vert 2\rangle$, and the fourth
to $\vert 3\rangle$. }
\label{eeh}
\end{figure}

\begin{figure}
\begin{minipage}{0.5\textwidth}
\includegraphics[width=0.3\textwidth]{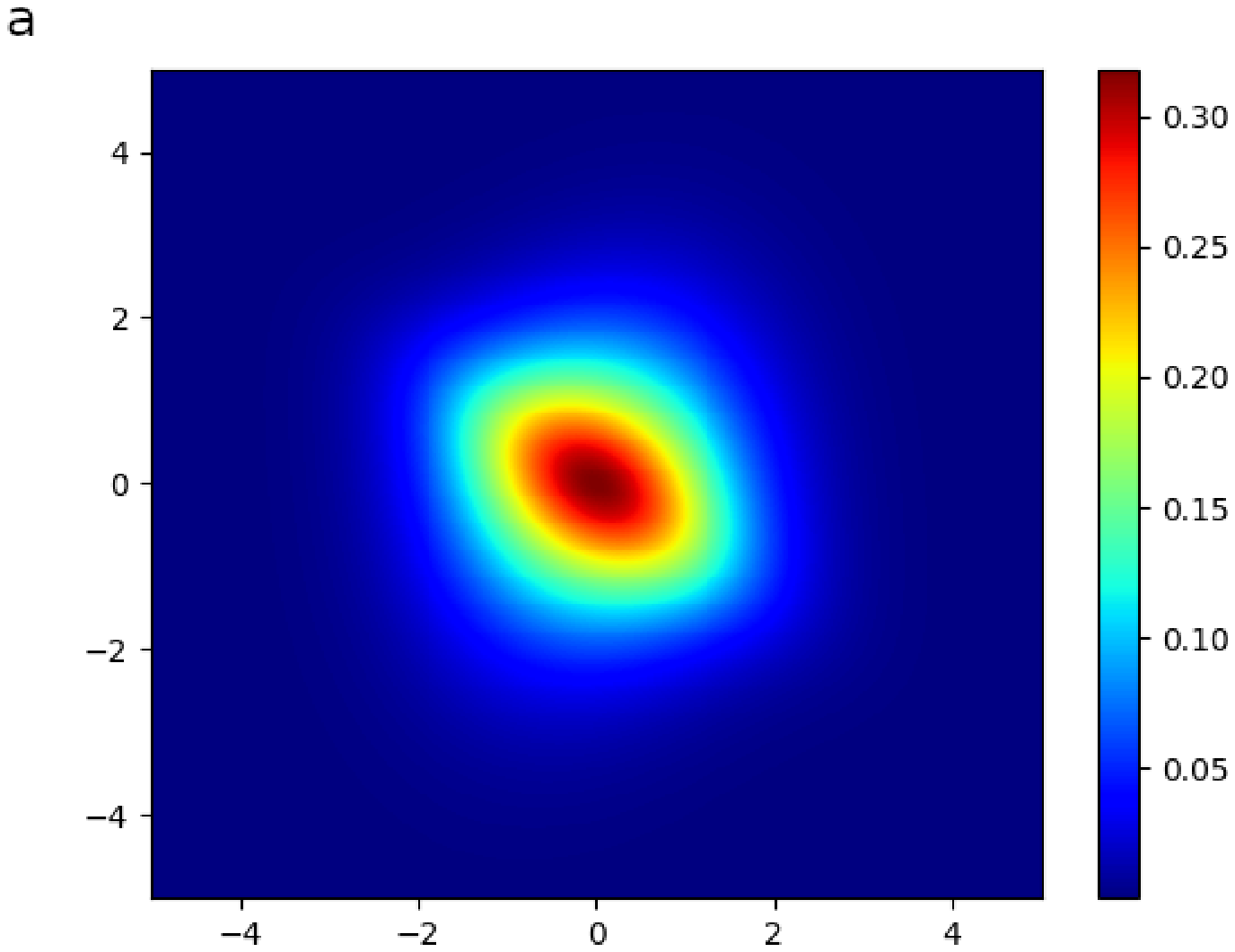}
\includegraphics[width=0.3\textwidth]{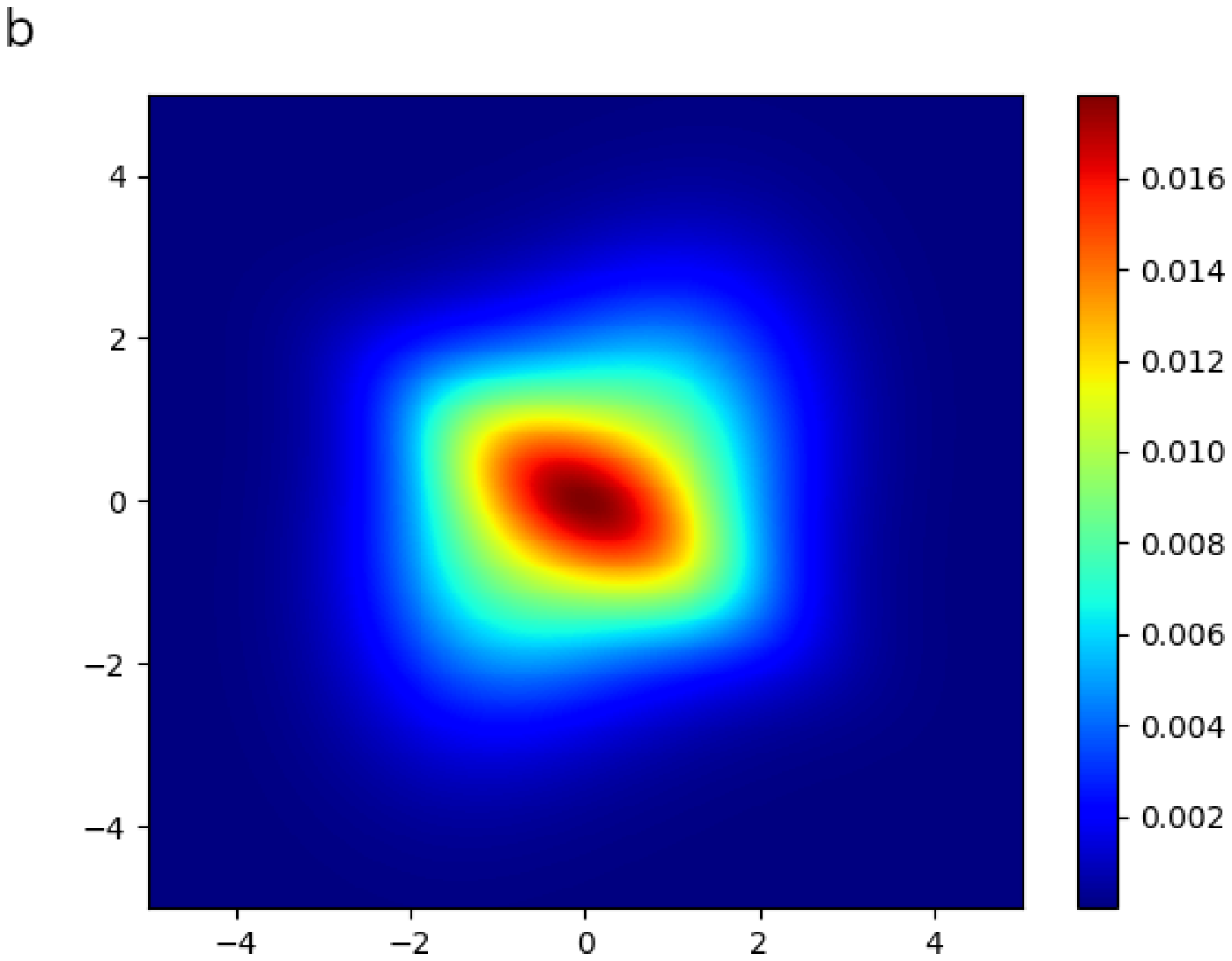}
\includegraphics[width=0.3\textwidth]{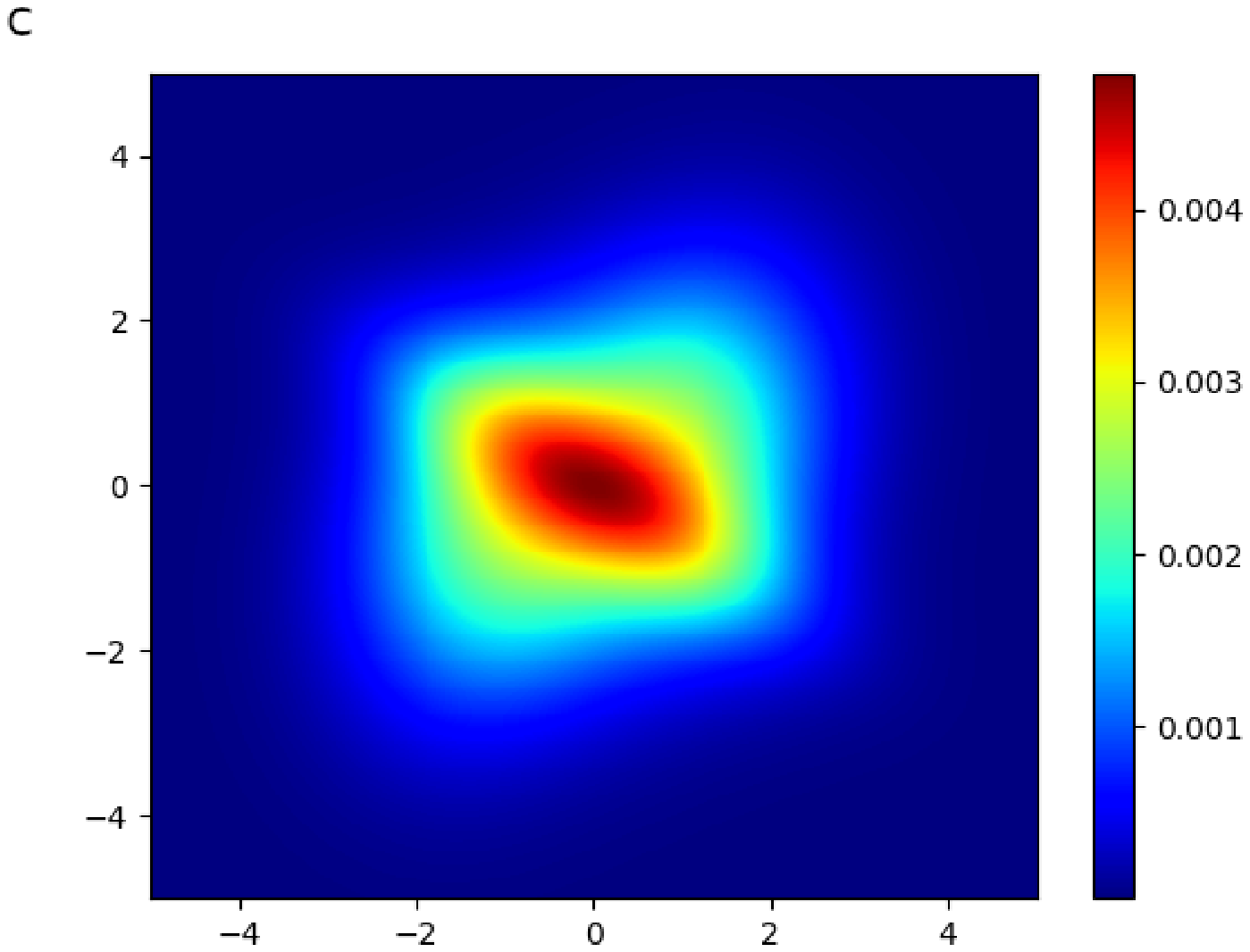}\\\quad
\includegraphics[width=0.3\textwidth]{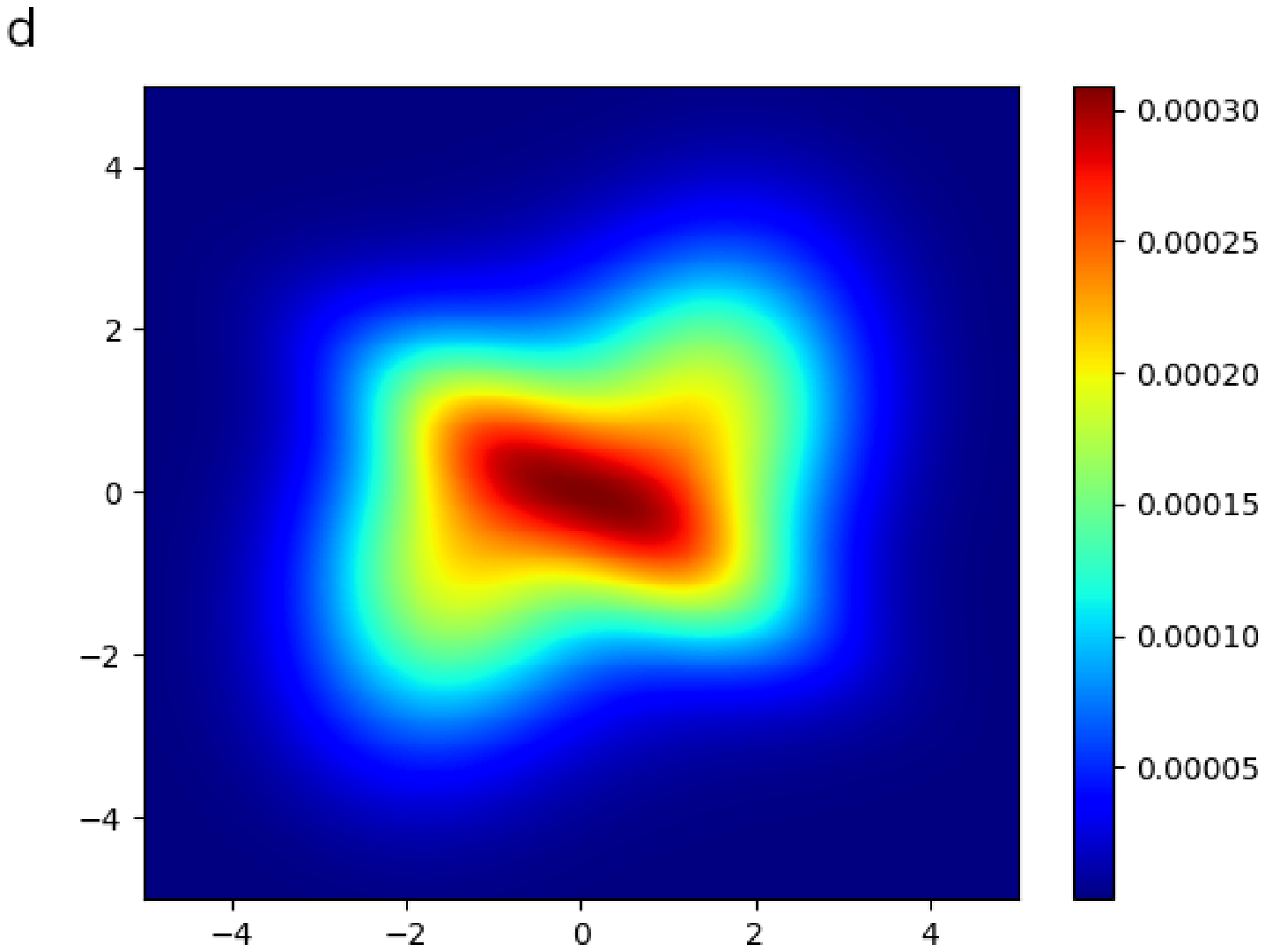}
\includegraphics[width=0.3\textwidth]{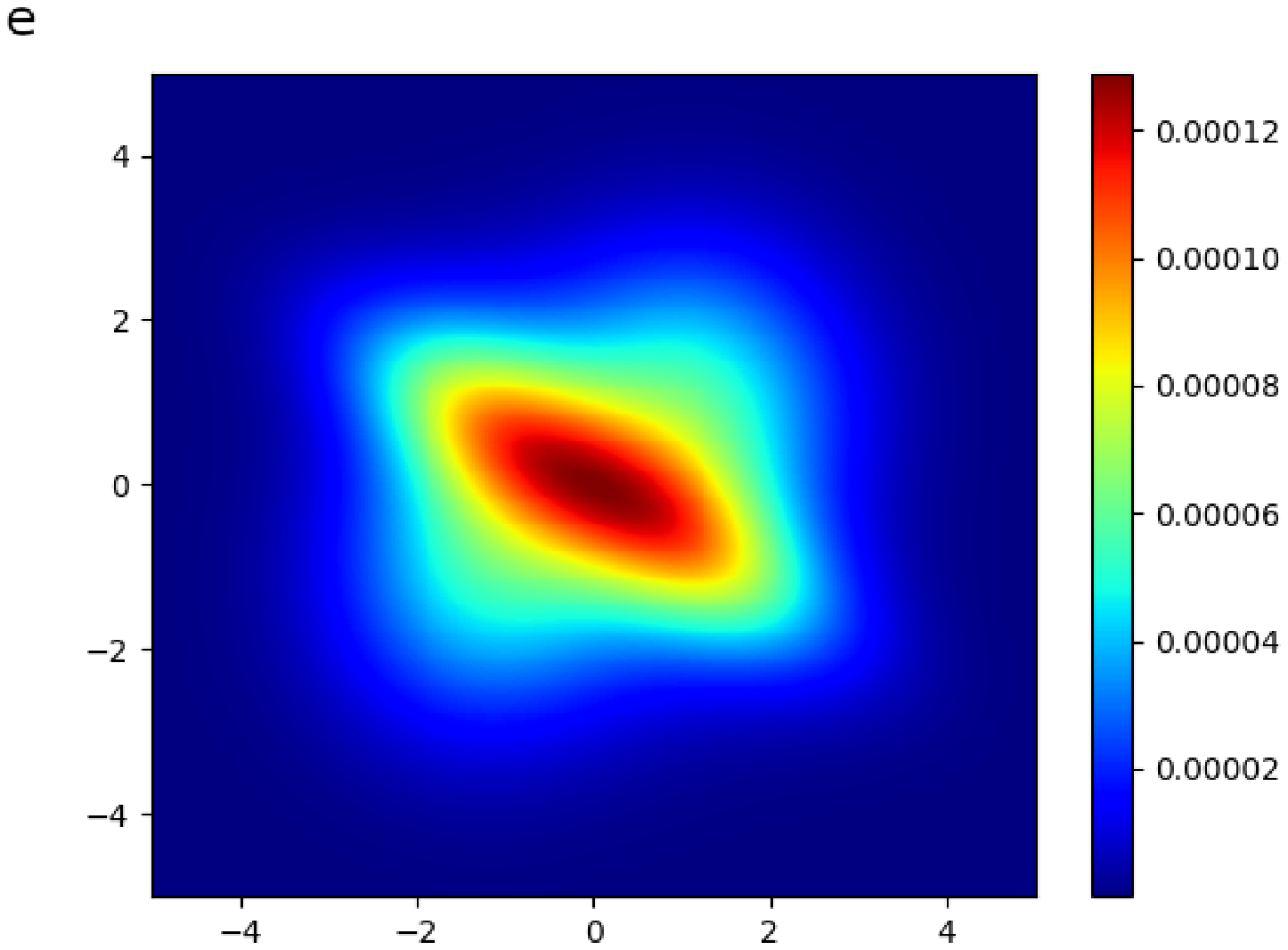}
\includegraphics[width=0.3\textwidth]{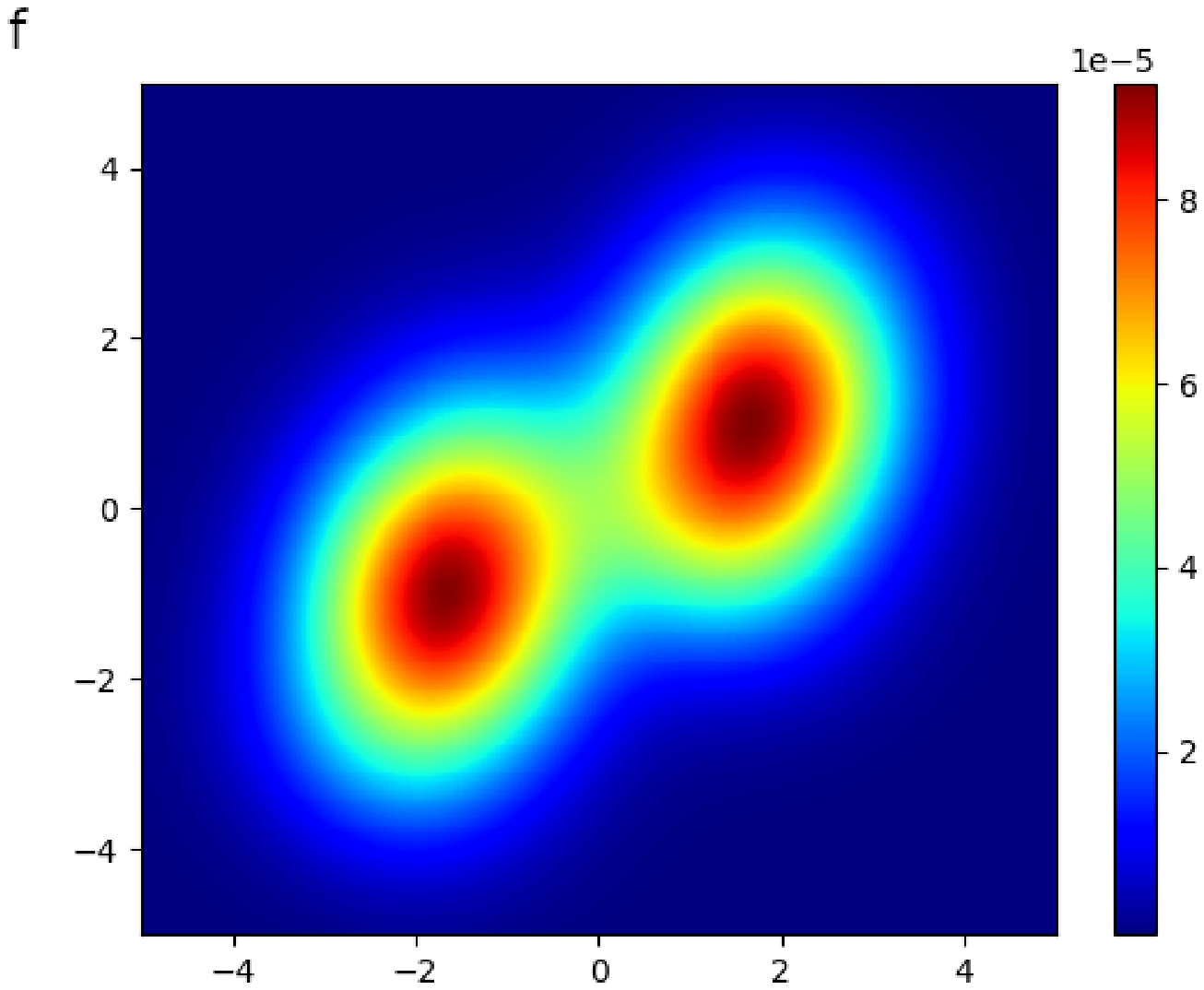}\\\quad
\end{minipage}
\caption{Two electrons and a hole in 2D coupled to two photon modes
($\omega_1=2,\omega_2=4$,$\vec{\lambda}_1=(4,4)$,$\vec{\lambda}_2=(-4,-4)$
a.u.) The total single particle  densities in photon space 
$|0\rangle|0\rangle$ (a), 
$|1\rangle|0\rangle$ (b), 
$|1\rangle|1\rangle$ (c), 
$|2\rangle|2\rangle$ (d). 
The electron density in $|2\rangle|2\rangle$ (e)
and the hole density in $|2\rangle|2\rangle$ (f) are also shown. 
}
\label{eeh2p}
\end{figure}

The next example is a trion, two-electron and the hole in 2D coupled
to one-photon mode (parameters are in the caption of Fig. \ref{eeh})
and two-photon
modes (Fig. \ref{eeh2p} lists the parameters).
Trions in optical microcavities in 2D semiconducting transition metal dichalcogenides 
have received intense experimental interest \cite{Emmanuele2020,Dhara2018, 
Dufferwiel2017,Sidler2017} due to the opportunities to engineer nanoscale
light–matter interactions. The calculated electron densities of the trion coupled
to one and two-photon modes are shown in Figs. \ref{eeh} and
\ref{eeh2p}. The figures show the density
\begin{equation}
\rho_{\vec{n}}(\vec{r})=\sum_{i=1}^3\sum_{j,k=1}^{K_{\vec{n}}}
c_j^{\vec{n}}
c_k^{\vec{n}} \langle
\psi_j^{\vec{n}}\vert\delta(\mathbf{r}_i-\mathbf{r})\vert
\psi_k^{\vec{n}}\rangle
\end{equation}
belonging to photon state $\vec{n}$ and we will also separate the
contribution of electrons and hole to the density. 

Figs. \ref{eeh}a and \ref{eeh2p}b show the density in the
$\vert 0\rangle$ and the  $\vert 0\rangle\vert 0\rangle$ photon spaces.
The total densities are very similar to these densities
because the $\vert 0\rangle$ and $\vert 0\rangle\vert 0\rangle$ 
photon spaces are dominant (92 and 82 percent of the wave
function belongs to these states). The weights of wave functions
components in photon spaces depend on the frequency and coupling, and by changing
these, other photon states (or linear
combinations of photon states) become major contributors to the
density. The total densities are very similar to each other in 
the one and two-photon mode coupled cases (Figs. \ref{eeh}a and
\ref{eeh2p}b): they are slightly
squeezed along the $x=-y$ diagonal axis. The diagonal symmetry is due to the
choice of the polarization vectors. 

Figs. \ref{eeh}b and \ref{eeh}c show the electron and hole density of
the trion in the $\vert 0 \rangle$ photon space. 
The hole density is circularly symmetric,
the electron density is elongated and the superposition of these two explains the structure seen in
Fig. \ref{eeh}a. In the
$\vert 1\rangle$ photon space the density is elongated (Fig. \ref{eeh}d) 
but now it is aligned in the direction of the $x=y$ diagonal. This
density is a superposition of two outside electron peaks (Fig. \ref{eeh}e) 
and a hole in the middle (Fig. \ref{eeh}f). The $\vert 2\rangle$ photon space is similar
(Fig. \ref{eeh}g) but now the electrons and the hole changed roles:
the electron density is in the middle and the hole density has two peaks
outside along the diagonal (Figs. \ref{eeh}h and \ref{eeh}i). The
electron density not only has a peak in the middle but also a bond-like structure
toward the hole peaks. The $\vert 3\rangle$ photon space densities 
(Figs. \ref{eeh}j, \ref{eeh}k, and \ref{eeh}l) are similar to the
$\vert 0 \rangle$ ones, but now
the total density has three separate peaks. 

In the trion coupled to  the two-photon  mode case (Fig. \ref{eeh2p}) 
we keep $\omega_1$ and ${\vec{\lambda}}_1$ the same as before and 
add a second photon mode with
$\omega_2=2\omega_1$ and ${\vec{\lambda}}_2$ =\,- ${\vec{\lambda}}_1$.
The second mode has stronger coupling because the frequency is larger. 
We still see  gradual elongation in higher photon spaces, but the
densities are quite different from the one-photon mode case. Now the
photon modes elongate the density along the diagonal in certain 
photon spaces and perpendicular to it in other spaces. The
resulted linear combinations are shown in Figs. 
\ref{eeh2p}b, \ref{eeh2p}c and \ref{eeh2p}d. The underlying electron
and hole densities still show somewhat similar structures to the 
one-photon mode. For example, the total density in the $\vert 2\rangle
\vert 2 \rangle$ photon space (Fig. \ref{eeh2p}d) is the sum of the 
electron density  in the  middle (Fig. \ref{eeh2p}e) and the two peaks
of the hole (Fig \ref{eeh2p}f) outside.  

\begin{figure}
\includegraphics[width=0.45\textwidth]{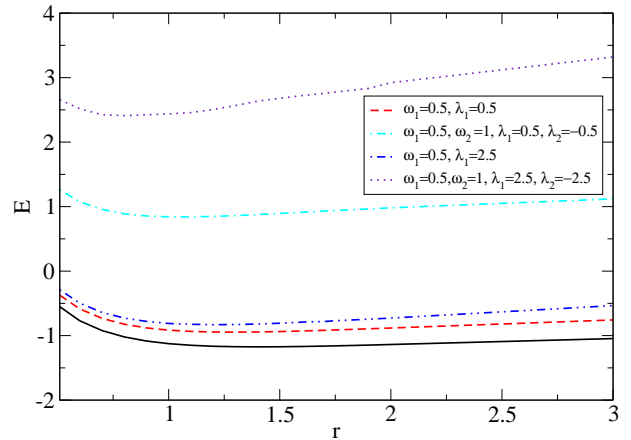}
\caption{Ground state energy of H$_2$ as a function of bond length (solid line)
and energies of the H$_2$ molecule coupled to light.
${\vec{\lambda}}_i=\lambda_i(1,1,0)$.}
\label{H2}
\end{figure}

Next, we study the energy as a function of the distance between the
nuclei in a 3-dimensional H$_2$ molecule coupled to one and two-photon modes.
This system was studied using one-dimensional model Hamiltonian in
Refs. \cite{doi:10.1063/5.0012723,doi:10.1021/acsphotonics.9b00648}
to explore the role of multiphoton modes and to build better approaches
to describe molecules in cavities. In this case, two protons are fixed
at a distance $r$ from each other and the energy is calculated as a
function of $r$. We have studied four cases, the 4 energy curves, 
together with the ground state energy curve are
shown in Fig. \ref{H2}. The energy of the system shifts up by coupling
to light, and the energy shifts higher when the ${\vec{\lambda}}$
coupling is larger. The effect of $\omega$ (not shown in the figure)
is much smaller. The shift in the two-photon mode case is much larger
because now two dipole self-interaction terms increase the energy. The
most interesting result is that by increasing ${\vec{\lambda}}$
the energy minimum moves to much shorter distances. In the ground
state the energy minimum is at $r=1.4$ a.u., while in the two-photon mode 
the equilibrium H$_2$ bond length in the present case is (dotted line in
Fig. \ref{H2})  $r=0.8$ a.u.

\begin{figure}
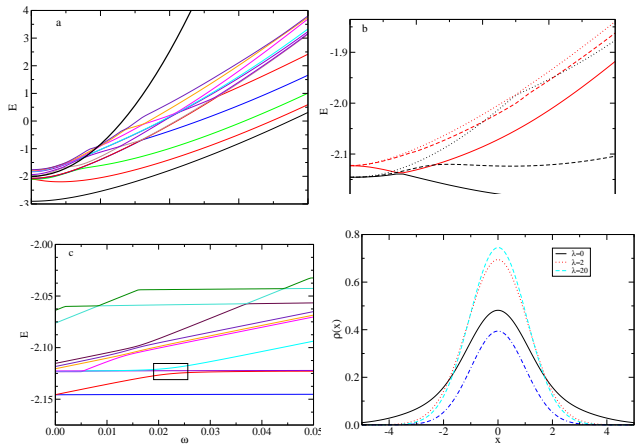

\begin{minipage}{0.5\textwidth}
\includegraphics[width=0.45\textwidth]{figure6a.eps}
\includegraphics[width=0.45\textwidth]{figure6b.eps}\\\quad
\includegraphics[width=0.45\textwidth]{figure6c.eps}
\includegraphics[width=0.45\textwidth]{figure6d.eps}\\\quad
\end{minipage}
\caption{(a) Energy of the lowest lying states of He as a function of
$\lambda$ for $\omega=0.8$ a.u. (b) Energy of the 2$^1S^e$ (starts at
-2.14 a.u.) and 2$^1P^o$ (starts at -2.12 a.u.) as a function of
$\lambda$, for $\omega=0.8$ a.u. (solid lines), $\omega=0.4$ a.u.
(dashed lines) and $\omega=0.2$ a.u. (dot-dashed line). (c) Energy of
states lying around the 2$^1S^e$ and 2$^1P^o$ as a function of $\omega$ 
($\sqrt{\omega}/\lambda$ is kept constant). (d) Particle density of the He atom
for different $\lambda$ values ($\omega=0.8$ a.u.)}
\label{He}
\end{figure}

The last example is a spin singlet He atom with finite nuclear mass coupled to a
single photon mode. Fig. \ref{He}a shows the lowest energy levels as a function of
${\vec{\lambda}}=(\lambda, \lambda,0)$  for $\omega=0.8$ a.u. Besides the bound
$S$,$P$, and $D$ states of the He atom, discretized continuum states
(finite box approximation of the continuum states) are also included for $\lambda=0$. 
The thick line shows the energy of the dissociation threshold into He$^+$ ion
plus an electron. Fig. \ref{He}a shows
that even some continuum states will become bound as $\lambda$
increases. The figure also shows that certain energy levels get  very
close to each other and then they move away (avoided crossing). This
energy level repulsion is due to the fact that changing $\lambda$
modifies the potential shape which, in turn, drives up or down energy levels
but no degeneracy is introduced and the energy levels cannot cross
each other. 

The role of $\omega$ is illustrated (see Fig. \ref{He}b) for the 2 $^1S^e$
and 2 $^1P^o$ energy levels. For $\omega$=0.8 a.u., the 2 $^1P^o$ state
moves down, approaching the  2 $^1S^e$ level at around $\lambda=0.08$
a.u., and then it moves up while the energy of the other state decreases 
(and later increases as Fig \ref{He}a shows). In the case of $\omega=0.4$
a.u. the 2 $^1S^e$ state moves up but the closest distance between the
two states (at around $\lambda=0.12$ a.u.) is much larger than the
size of the  previous gap. In the  $\omega=0.2$ a.u. case the two curves 
meet at $\lambda=0.27$
a.u. These examples  show that $\omega$ strongly influences 
the position of the avoided crossing points and it also affects rise of 
the energy curves. 

Fig. \ref{He}c shows the energy levels at even lower coupling strength. 
In this case we have changed $\omega$ keeping $\sqrt{\omega}/\lambda=23.45$
fixed. In this way at $\omega=E_{2 ^1S^e}-E_{2 ^1P^o}=0.22$ a.u.
$\lambda$ is 0.02 a.u. and the coupling strength is proportional to
$\omega$ as suggested in Ref. \cite{doi:10.1021/acs.jpclett.0c01556}
to mimic the Jaynes-Cummings model \cite{1443594}. The
box in the middle of Fig. \ref{He}c shows the position where the 2 $^1S^e$ and 2 $^1P^o$
state gets close to each other at around the $\omega=0.22$ a.u.
transition energy and at that point
the energy difference between them is 0.0055 a.u., which
corresponds to a Rabi splitting of 0.1496 eV in excellent agreement
with the real space grid based calculation (0.148 eV) in Ref.
\cite{doi:10.1021/acs.jpclett.0c01556}. 

Finally, Fig. \ref{He}d shows the particle density of the He atom for
different couplings. The figure shows the average density in the $x$
direction, and averaging in the $y$ or $z$ direction gives the same curve
as the average in the $x$ direction. The lowest curve is the density
distribution  of the He nucleus
and it remains the same for different $\lambda$ values. Increasing 
$\lambda$ squeezes the density toward the middle, and the system becomes
more compact, but the difference between the densities 
for $\lambda$=2 a.u. and  for $\lambda$=20 a.u. is small. This is
probably because increasing $\lambda$ couples higher 
photon spaces to the wave function and in these spaces the wave
function components have different parity and spatial extension, so the effect 
of increasing $\lambda$ is not just a simple squeezing of the density.

In summary, we have developed an approach to accurately calculate the
light-matter coupled wave functions. The presented examples show that
the hybrid states are very different from the original electronic states
and the wave function has to be optimized in every photon space to
accurately represent the system. Our approach works for charged and
neutral systems, and the center of mass motion can be fixed or removed.
The ECGs are limited to up to 8-10 particles due to the $N_i!$ 
scaling of the explicit antisymmetrization ($N_i$ is the number of identical 
particles in the system), but other forms of trial functions, e.g.
configuration interaction type functions, might be also tested.

\begin{acknowledgments}
This work has been supported by the National Science
Foundation (NSF) under Grant No. IRES 1826917. AA, CH and MB equally contributed to this work. 
\end{acknowledgments}


\begin{thebibliography}{73}
\expandafter\ifx\csname natexlab\endcsname\relax\def\natexlab#1{#1}\fi
\expandafter\ifx\csname bibnamefont\endcsname\relax
  \def\bibnamefont#1{#1}\fi
\expandafter\ifx\csname bibfnamefont\endcsname\relax
  \def\bibfnamefont#1{#1}\fi
\expandafter\ifx\csname citenamefont\endcsname\relax
  \def\citenamefont#1{#1}\fi
\expandafter\ifx\csname url\endcsname\relax
  \def\url#1{\texttt{#1}}\fi
\expandafter\ifx\csname urlprefix\endcsname\relax\def\urlprefix{URL }\fi
\providecommand{\bibinfo}[2]{#2}
\providecommand{\eprint}[2][]{\url{#2}}

\bibitem[{\citenamefont{Feist and Garcia-Vidal}(2015)}]{PhysRevLett.114.196402}
\bibinfo{author}{\bibfnamefont{J.}~\bibnamefont{Feist}} \bibnamefont{and}
  \bibinfo{author}{\bibfnamefont{F.~J.} \bibnamefont{Garcia-Vidal}},
  \bibinfo{journal}{Phys. Rev. Lett.} \textbf{\bibinfo{volume}{114}},
  \bibinfo{pages}{196402} (\bibinfo{year}{2015}),
  \urlprefix\url{https://link.aps.org/doi/10.1103/PhysRevLett.114.196402}.

\bibitem[{\citenamefont{Balili et~al.}(2007)\citenamefont{Balili, Hartwell,
  Snoke, Pfeiffer, and West}}]{Balili1007}
\bibinfo{author}{\bibfnamefont{R.}~\bibnamefont{Balili}},
  \bibinfo{author}{\bibfnamefont{V.}~\bibnamefont{Hartwell}},
  \bibinfo{author}{\bibfnamefont{D.}~\bibnamefont{Snoke}},
  \bibinfo{author}{\bibfnamefont{L.}~\bibnamefont{Pfeiffer}}, \bibnamefont{and}
  \bibinfo{author}{\bibfnamefont{K.}~\bibnamefont{West}},
  \bibinfo{journal}{Science} \textbf{\bibinfo{volume}{316}},
  \bibinfo{pages}{1007} (\bibinfo{year}{2007}), ISSN \bibinfo{issn}{0036-8075},
  \eprint{https://science.sciencemag.org/content/316/5827/1007.full.pdf},
  \urlprefix\url{https://science.sciencemag.org/content/316/5827/1007}.

\bibitem[{\citenamefont{Schachenmayer et~al.}(2015)\citenamefont{Schachenmayer,
  Genes, Tignone, and Pupillo}}]{PhysRevLett.114.196403}
\bibinfo{author}{\bibfnamefont{J.}~\bibnamefont{Schachenmayer}},
  \bibinfo{author}{\bibfnamefont{C.}~\bibnamefont{Genes}},
  \bibinfo{author}{\bibfnamefont{E.}~\bibnamefont{Tignone}}, \bibnamefont{and}
  \bibinfo{author}{\bibfnamefont{G.}~\bibnamefont{Pupillo}},
  \bibinfo{journal}{Phys. Rev. Lett.} \textbf{\bibinfo{volume}{114}},
  \bibinfo{pages}{196403} (\bibinfo{year}{2015}),
  \urlprefix\url{https://link.aps.org/doi/10.1103/PhysRevLett.114.196403}.

\bibitem[{\citenamefont{Xiang et~al.}(2020)\citenamefont{Xiang, Ribeiro, Du,
  Chen, Yang, Wang, Yuen-Zhou, and Xiong}}]{Xiang665}
\bibinfo{author}{\bibfnamefont{B.}~\bibnamefont{Xiang}},
  \bibinfo{author}{\bibfnamefont{R.~F.} \bibnamefont{Ribeiro}},
  \bibinfo{author}{\bibfnamefont{M.}~\bibnamefont{Du}},
  \bibinfo{author}{\bibfnamefont{L.}~\bibnamefont{Chen}},
  \bibinfo{author}{\bibfnamefont{Z.}~\bibnamefont{Yang}},
  \bibinfo{author}{\bibfnamefont{J.}~\bibnamefont{Wang}},
  \bibinfo{author}{\bibfnamefont{J.}~\bibnamefont{Yuen-Zhou}},
  \bibnamefont{and} \bibinfo{author}{\bibfnamefont{W.}~\bibnamefont{Xiong}},
  \bibinfo{journal}{Science} \textbf{\bibinfo{volume}{368}},
  \bibinfo{pages}{665} (\bibinfo{year}{2020}), ISSN \bibinfo{issn}{0036-8075},
  \eprint{https://science.sciencemag.org/content/368/6491/665.full.pdf},
  \urlprefix\url{https://science.sciencemag.org/content/368/6491/665}.

\bibitem[{\citenamefont{Reserbat-Plantey
  et~al.}(2021)\citenamefont{Reserbat-Plantey, Epstein, Torre, Costa,
  Gonçalves, Mortensen, Polini, Song, Peres, and
  Koppens}}]{doi:10.1021/acsphotonics.0c01224}
\bibinfo{author}{\bibfnamefont{A.}~\bibnamefont{Reserbat-Plantey}},
  \bibinfo{author}{\bibfnamefont{I.}~\bibnamefont{Epstein}},
  \bibinfo{author}{\bibfnamefont{I.}~\bibnamefont{Torre}},
  \bibinfo{author}{\bibfnamefont{A.~T.} \bibnamefont{Costa}},
  \bibinfo{author}{\bibfnamefont{P.~A.~D.} \bibnamefont{Gonçalves}},
  \bibinfo{author}{\bibfnamefont{N.~A.} \bibnamefont{Mortensen}},
  \bibinfo{author}{\bibfnamefont{M.}~\bibnamefont{Polini}},
  \bibinfo{author}{\bibfnamefont{J.~C.~W.} \bibnamefont{Song}},
  \bibinfo{author}{\bibfnamefont{N.~M.~R.} \bibnamefont{Peres}},
  \bibnamefont{and} \bibinfo{author}{\bibfnamefont{F.~H.~L.}
  \bibnamefont{Koppens}}, \bibinfo{journal}{ACS Photonics}
  \textbf{\bibinfo{volume}{8}}, \bibinfo{pages}{85} (\bibinfo{year}{2021}),
  \urlprefix\url{https://doi.org/10.1021/acsphotonics.0c01224}.

\bibitem[{\citenamefont{Coles et~al.}(2014)\citenamefont{Coles, Somaschi,
  Michetti, Clark, Lagoudakis, Savvidis, and Lidzey}}]{Coles2014}
\bibinfo{author}{\bibfnamefont{D.~M.} \bibnamefont{Coles}},
  \bibinfo{author}{\bibfnamefont{N.}~\bibnamefont{Somaschi}},
  \bibinfo{author}{\bibfnamefont{P.}~\bibnamefont{Michetti}},
  \bibinfo{author}{\bibfnamefont{C.}~\bibnamefont{Clark}},
  \bibinfo{author}{\bibfnamefont{P.~G.} \bibnamefont{Lagoudakis}},
  \bibinfo{author}{\bibfnamefont{P.~G.} \bibnamefont{Savvidis}},
  \bibnamefont{and} \bibinfo{author}{\bibfnamefont{D.~G.}
  \bibnamefont{Lidzey}}, \bibinfo{journal}{Nature Materials}
  \textbf{\bibinfo{volume}{13}}, \bibinfo{pages}{712} (\bibinfo{year}{2014}),
  ISSN \bibinfo{issn}{1476-4660},
  \urlprefix\url{https://doi.org/10.1038/nmat3950}.

\bibitem[{\citenamefont{Kasprzak et~al.}(2006)\citenamefont{Kasprzak, Richard,
  Kundermann, Baas, Jeambrun, Keeling, Marchetti, Szyma{\'{n}}ska, Andr{\'e},
  Staehli et~al.}}]{Kasprzak2006}
\bibinfo{author}{\bibfnamefont{J.}~\bibnamefont{Kasprzak}},
  \bibinfo{author}{\bibfnamefont{M.}~\bibnamefont{Richard}},
  \bibinfo{author}{\bibfnamefont{S.}~\bibnamefont{Kundermann}},
  \bibinfo{author}{\bibfnamefont{A.}~\bibnamefont{Baas}},
  \bibinfo{author}{\bibfnamefont{P.}~\bibnamefont{Jeambrun}},
  \bibinfo{author}{\bibfnamefont{J.~M.~J.} \bibnamefont{Keeling}},
  \bibinfo{author}{\bibfnamefont{F.~M.} \bibnamefont{Marchetti}},
  \bibinfo{author}{\bibfnamefont{M.~H.} \bibnamefont{Szyma{\'{n}}ska}},
  \bibinfo{author}{\bibfnamefont{R.}~\bibnamefont{Andr{\'e}}},
  \bibinfo{author}{\bibfnamefont{J.~L.} \bibnamefont{Staehli}},
  \bibnamefont{et~al.}, \bibinfo{journal}{Nature}
  \textbf{\bibinfo{volume}{443}}, \bibinfo{pages}{409} (\bibinfo{year}{2006}),
  ISSN \bibinfo{issn}{1476-4687},
  \urlprefix\url{https://doi.org/10.1038/nature05131}.

\bibitem[{\citenamefont{Schwartz et~al.}(2011)\citenamefont{Schwartz,
  Hutchison, Genet, and Ebbesen}}]{PhysRevLett.106.196405}
\bibinfo{author}{\bibfnamefont{T.}~\bibnamefont{Schwartz}},
  \bibinfo{author}{\bibfnamefont{J.~A.} \bibnamefont{Hutchison}},
  \bibinfo{author}{\bibfnamefont{C.}~\bibnamefont{Genet}}, \bibnamefont{and}
  \bibinfo{author}{\bibfnamefont{T.~W.} \bibnamefont{Ebbesen}},
  \bibinfo{journal}{Phys. Rev. Lett.} \textbf{\bibinfo{volume}{106}},
  \bibinfo{pages}{196405} (\bibinfo{year}{2011}),
  \urlprefix\url{https://link.aps.org/doi/10.1103/PhysRevLett.106.196405}.

\bibitem[{\citenamefont{Plumhof et~al.}(2014)\citenamefont{Plumhof,
  St{\"o}ferle, Mai, Scherf, and Mahrt}}]{Plumhof2014}
\bibinfo{author}{\bibfnamefont{J.~D.} \bibnamefont{Plumhof}},
  \bibinfo{author}{\bibfnamefont{T.}~\bibnamefont{St{\"o}ferle}},
  \bibinfo{author}{\bibfnamefont{L.}~\bibnamefont{Mai}},
  \bibinfo{author}{\bibfnamefont{U.}~\bibnamefont{Scherf}}, \bibnamefont{and}
  \bibinfo{author}{\bibfnamefont{R.~F.} \bibnamefont{Mahrt}},
  \bibinfo{journal}{Nature Materials} \textbf{\bibinfo{volume}{13}},
  \bibinfo{pages}{247} (\bibinfo{year}{2014}), ISSN \bibinfo{issn}{1476-4660},
  \urlprefix\url{https://doi.org/10.1038/nmat3825}.

\bibitem[{\citenamefont{Hutchison et~al.}(2013)\citenamefont{Hutchison, Liscio,
  Schwartz, Canaguier-Durand, Genet, Palermo, Samorì, and
  Ebbesen}}]{https://doi.org/10.1002/adma.201203682}
\bibinfo{author}{\bibfnamefont{J.~A.} \bibnamefont{Hutchison}},
  \bibinfo{author}{\bibfnamefont{A.}~\bibnamefont{Liscio}},
  \bibinfo{author}{\bibfnamefont{T.}~\bibnamefont{Schwartz}},
  \bibinfo{author}{\bibfnamefont{A.}~\bibnamefont{Canaguier-Durand}},
  \bibinfo{author}{\bibfnamefont{C.}~\bibnamefont{Genet}},
  \bibinfo{author}{\bibfnamefont{V.}~\bibnamefont{Palermo}},
  \bibinfo{author}{\bibfnamefont{P.}~\bibnamefont{Samorì}}, \bibnamefont{and}
  \bibinfo{author}{\bibfnamefont{T.~W.} \bibnamefont{Ebbesen}},
  \bibinfo{journal}{Advanced Materials} \textbf{\bibinfo{volume}{25}},
  \bibinfo{pages}{2481} (\bibinfo{year}{2013}),
  \eprint{https://onlinelibrary.wiley.com/doi/pdf/10.1002/adma.201203682},
  \urlprefix\url{https://onlinelibrary.wiley.com/doi/abs/10.1002/adma.201203682}.

\bibitem[{\citenamefont{Wang et~al.}(2021{\natexlab{a}})\citenamefont{Wang,
  Seidel, Nagarajan, Chervy, Genet, and Ebbesen}}]{Wang2021}
\bibinfo{author}{\bibfnamefont{K.}~\bibnamefont{Wang}},
  \bibinfo{author}{\bibfnamefont{M.}~\bibnamefont{Seidel}},
  \bibinfo{author}{\bibfnamefont{K.}~\bibnamefont{Nagarajan}},
  \bibinfo{author}{\bibfnamefont{T.}~\bibnamefont{Chervy}},
  \bibinfo{author}{\bibfnamefont{C.}~\bibnamefont{Genet}}, \bibnamefont{and}
  \bibinfo{author}{\bibfnamefont{T.}~\bibnamefont{Ebbesen}},
  \bibinfo{journal}{Nature Communications} \textbf{\bibinfo{volume}{12}},
  \bibinfo{pages}{1486} (\bibinfo{year}{2021}{\natexlab{a}}), ISSN
  \bibinfo{issn}{2041-1723},
  \urlprefix\url{https://doi.org/10.1038/s41467-021-21739-7}.

\bibitem[{\citenamefont{Basov et~al.}(2021)\citenamefont{Basov, Asenjo-Garcia,
  Schuck, Zhu, and Rubio}}]{BasovAsenjoGarciaSchuckZhuRubio+2021+549+577}
\bibinfo{author}{\bibfnamefont{D.~N.} \bibnamefont{Basov}},
  \bibinfo{author}{\bibfnamefont{A.}~\bibnamefont{Asenjo-Garcia}},
  \bibinfo{author}{\bibfnamefont{P.~J.} \bibnamefont{Schuck}},
  \bibinfo{author}{\bibfnamefont{X.}~\bibnamefont{Zhu}}, \bibnamefont{and}
  \bibinfo{author}{\bibfnamefont{A.}~\bibnamefont{Rubio}},
  \bibinfo{journal}{Nanophotonics} \textbf{\bibinfo{volume}{10}},
  \bibinfo{pages}{549} (\bibinfo{year}{2021}),
  \urlprefix\url{https://doi.org/10.1515/nanoph-2020-0449}.

\bibitem[{\citenamefont{Mazza and Georges}(2019)}]{PhysRevLett.122.017401}
\bibinfo{author}{\bibfnamefont{G.}~\bibnamefont{Mazza}} \bibnamefont{and}
  \bibinfo{author}{\bibfnamefont{A.}~\bibnamefont{Georges}},
  \bibinfo{journal}{Phys. Rev. Lett.} \textbf{\bibinfo{volume}{122}},
  \bibinfo{pages}{017401} (\bibinfo{year}{2019}),
  \urlprefix\url{https://link.aps.org/doi/10.1103/PhysRevLett.122.017401}.

\bibitem[{\citenamefont{Di~Stefano et~al.}(2019)\citenamefont{Di~Stefano,
  Settineri, Macr{\`i}, Garziano, Stassi, Savasta, and Nori}}]{DiStefano2019}
\bibinfo{author}{\bibfnamefont{O.}~\bibnamefont{Di~Stefano}},
  \bibinfo{author}{\bibfnamefont{A.}~\bibnamefont{Settineri}},
  \bibinfo{author}{\bibfnamefont{V.}~\bibnamefont{Macr{\`i}}},
  \bibinfo{author}{\bibfnamefont{L.}~\bibnamefont{Garziano}},
  \bibinfo{author}{\bibfnamefont{R.}~\bibnamefont{Stassi}},
  \bibinfo{author}{\bibfnamefont{S.}~\bibnamefont{Savasta}}, \bibnamefont{and}
  \bibinfo{author}{\bibfnamefont{F.}~\bibnamefont{Nori}},
  \bibinfo{journal}{Nature Physics} \textbf{\bibinfo{volume}{15}},
  \bibinfo{pages}{803} (\bibinfo{year}{2019}), ISSN \bibinfo{issn}{1745-2481},
  \urlprefix\url{https://doi.org/10.1038/s41567-019-0534-4}.

\bibitem[{\citenamefont{Galego et~al.}(2015)\citenamefont{Galego, Garcia-Vidal,
  and Feist}}]{PhysRevX.5.041022}
\bibinfo{author}{\bibfnamefont{J.}~\bibnamefont{Galego}},
  \bibinfo{author}{\bibfnamefont{F.~J.} \bibnamefont{Garcia-Vidal}},
  \bibnamefont{and} \bibinfo{author}{\bibfnamefont{J.}~\bibnamefont{Feist}},
  \bibinfo{journal}{Phys. Rev. X} \textbf{\bibinfo{volume}{5}},
  \bibinfo{pages}{041022} (\bibinfo{year}{2015}),
  \urlprefix\url{https://link.aps.org/doi/10.1103/PhysRevX.5.041022}.

\bibitem[{\citenamefont{Herrera and Spano}(2016)}]{PhysRevLett.116.238301}
\bibinfo{author}{\bibfnamefont{F.}~\bibnamefont{Herrera}} \bibnamefont{and}
  \bibinfo{author}{\bibfnamefont{F.~C.} \bibnamefont{Spano}},
  \bibinfo{journal}{Phys. Rev. Lett.} \textbf{\bibinfo{volume}{116}},
  \bibinfo{pages}{238301} (\bibinfo{year}{2016}),
  \urlprefix\url{https://link.aps.org/doi/10.1103/PhysRevLett.116.238301}.

\bibitem[{\citenamefont{Galego et~al.}(2016)\citenamefont{Galego, Garcia-Vidal,
  and Feist}}]{Galego2016}
\bibinfo{author}{\bibfnamefont{J.}~\bibnamefont{Galego}},
  \bibinfo{author}{\bibfnamefont{F.~J.} \bibnamefont{Garcia-Vidal}},
  \bibnamefont{and} \bibinfo{author}{\bibfnamefont{J.}~\bibnamefont{Feist}},
  \bibinfo{journal}{Nature Communications} \textbf{\bibinfo{volume}{7}},
  \bibinfo{pages}{13841} (\bibinfo{year}{2016}), ISSN
  \bibinfo{issn}{2041-1723},
  \urlprefix\url{https://doi.org/10.1038/ncomms13841}.

\bibitem[{\citenamefont{Shalabney et~al.}(2015)\citenamefont{Shalabney, George,
  Hutchison, Pupillo, Genet, and Ebbesen}}]{Shalabney2015}
\bibinfo{author}{\bibfnamefont{A.}~\bibnamefont{Shalabney}},
  \bibinfo{author}{\bibfnamefont{J.}~\bibnamefont{George}},
  \bibinfo{author}{\bibfnamefont{J.}~\bibnamefont{Hutchison}},
  \bibinfo{author}{\bibfnamefont{G.}~\bibnamefont{Pupillo}},
  \bibinfo{author}{\bibfnamefont{C.}~\bibnamefont{Genet}}, \bibnamefont{and}
  \bibinfo{author}{\bibfnamefont{T.~W.} \bibnamefont{Ebbesen}},
  \bibinfo{journal}{Nature Communications} \textbf{\bibinfo{volume}{6}},
  \bibinfo{pages}{5981} (\bibinfo{year}{2015}), ISSN \bibinfo{issn}{2041-1723},
  \urlprefix\url{https://doi.org/10.1038/ncomms6981}.

\bibitem[{\citenamefont{Buchholz et~al.}(2019)\citenamefont{Buchholz,
  Theophilou, Nielsen, Ruggenthaler, and
  Rubio}}]{doi:10.1021/acsphotonics.9b00648}
\bibinfo{author}{\bibfnamefont{F.}~\bibnamefont{Buchholz}},
  \bibinfo{author}{\bibfnamefont{I.}~\bibnamefont{Theophilou}},
  \bibinfo{author}{\bibfnamefont{S.~E.~B.} \bibnamefont{Nielsen}},
  \bibinfo{author}{\bibfnamefont{M.}~\bibnamefont{Ruggenthaler}},
  \bibnamefont{and} \bibinfo{author}{\bibfnamefont{A.}~\bibnamefont{Rubio}},
  \bibinfo{journal}{ACS Photonics} \textbf{\bibinfo{volume}{6}},
  \bibinfo{pages}{2694} (\bibinfo{year}{2019}).

\bibitem[{\citenamefont{Sch{\"a}fer et~al.}(2019)\citenamefont{Sch{\"a}fer,
  Ruggenthaler, Appel, and Rubio}}]{Schafer4883}
\bibinfo{author}{\bibfnamefont{C.}~\bibnamefont{Sch{\"a}fer}},
  \bibinfo{author}{\bibfnamefont{M.}~\bibnamefont{Ruggenthaler}},
  \bibinfo{author}{\bibfnamefont{H.}~\bibnamefont{Appel}}, \bibnamefont{and}
  \bibinfo{author}{\bibfnamefont{A.}~\bibnamefont{Rubio}},
  \bibinfo{journal}{Proceedings of the National Academy of Sciences}
  \textbf{\bibinfo{volume}{116}}, \bibinfo{pages}{4883} (\bibinfo{year}{2019}),
  ISSN \bibinfo{issn}{0027-8424},
  \eprint{https://www.pnas.org/content/116/11/4883.full.pdf},
  \urlprefix\url{https://www.pnas.org/content/116/11/4883}.

\bibitem[{\citenamefont{Ruggenthaler et~al.}(2018)\citenamefont{Ruggenthaler,
  Tancogne-Dejean, Flick, Appel, and Rubio}}]{Ruggenthaler2018}
\bibinfo{author}{\bibfnamefont{M.}~\bibnamefont{Ruggenthaler}},
  \bibinfo{author}{\bibfnamefont{N.}~\bibnamefont{Tancogne-Dejean}},
  \bibinfo{author}{\bibfnamefont{J.}~\bibnamefont{Flick}},
  \bibinfo{author}{\bibfnamefont{H.}~\bibnamefont{Appel}}, \bibnamefont{and}
  \bibinfo{author}{\bibfnamefont{A.}~\bibnamefont{Rubio}},
  \bibinfo{journal}{Nature Reviews Chemistry} \textbf{\bibinfo{volume}{2}},
  \bibinfo{pages}{0118} (\bibinfo{year}{2018}), ISSN \bibinfo{issn}{2397-3358},
  \urlprefix\url{https://doi.org/10.1038/s41570-018-0118}.

\bibitem[{\citenamefont{Flick et~al.}(2015)\citenamefont{Flick, Ruggenthaler,
  Appel, and Rubio}}]{Flick15285}
\bibinfo{author}{\bibfnamefont{J.}~\bibnamefont{Flick}},
  \bibinfo{author}{\bibfnamefont{M.}~\bibnamefont{Ruggenthaler}},
  \bibinfo{author}{\bibfnamefont{H.}~\bibnamefont{Appel}}, \bibnamefont{and}
  \bibinfo{author}{\bibfnamefont{A.}~\bibnamefont{Rubio}},
  \bibinfo{journal}{Proceedings of the National Academy of Sciences}
  \textbf{\bibinfo{volume}{112}}, \bibinfo{pages}{15285}
  (\bibinfo{year}{2015}), ISSN \bibinfo{issn}{0027-8424},
  \eprint{https://www.pnas.org/content/112/50/15285.full.pdf},
  \urlprefix\url{https://www.pnas.org/content/112/50/15285}.

\bibitem[{\citenamefont{Flick et~al.}(2017)\citenamefont{Flick, Ruggenthaler,
  Appel, and Rubio}}]{Flick3026}
\bibinfo{author}{\bibfnamefont{J.}~\bibnamefont{Flick}},
  \bibinfo{author}{\bibfnamefont{M.}~\bibnamefont{Ruggenthaler}},
  \bibinfo{author}{\bibfnamefont{H.}~\bibnamefont{Appel}}, \bibnamefont{and}
  \bibinfo{author}{\bibfnamefont{A.}~\bibnamefont{Rubio}},
  \bibinfo{journal}{Proceedings of the National Academy of Sciences}
  \textbf{\bibinfo{volume}{114}}, \bibinfo{pages}{3026} (\bibinfo{year}{2017}),
  ISSN \bibinfo{issn}{0027-8424},
  \eprint{https://www.pnas.org/content/114/12/3026.full.pdf},
  \urlprefix\url{https://www.pnas.org/content/114/12/3026}.

\bibitem[{\citenamefont{Lacombe et~al.}(2019)\citenamefont{Lacombe, Hoffmann,
  and Maitra}}]{PhysRevLett.123.083201}
\bibinfo{author}{\bibfnamefont{L.}~\bibnamefont{Lacombe}},
  \bibinfo{author}{\bibfnamefont{N.~M.} \bibnamefont{Hoffmann}},
  \bibnamefont{and} \bibinfo{author}{\bibfnamefont{N.~T.}
  \bibnamefont{Maitra}}, \bibinfo{journal}{Phys. Rev. Lett.}
  \textbf{\bibinfo{volume}{123}}, \bibinfo{pages}{083201}
  (\bibinfo{year}{2019}),
  \urlprefix\url{https://link.aps.org/doi/10.1103/PhysRevLett.123.083201}.

\bibitem[{\citenamefont{Mandal et~al.}(2020{\natexlab{a}})\citenamefont{Mandal,
  Montillo~Vega, and Huo}}]{Mandal}
\bibinfo{author}{\bibfnamefont{A.}~\bibnamefont{Mandal}},
  \bibinfo{author}{\bibfnamefont{S.}~\bibnamefont{Montillo~Vega}},
  \bibnamefont{and} \bibinfo{author}{\bibfnamefont{P.}~\bibnamefont{Huo}},
  \bibinfo{journal}{The Journal of Physical Chemistry Letters}
  \textbf{\bibinfo{volume}{11}}, \bibinfo{pages}{9215}
  (\bibinfo{year}{2020}{\natexlab{a}}), \bibinfo{note}{pMID: 32991814}.

\bibitem[{\citenamefont{Flick et~al.}(2019)\citenamefont{Flick, Welakuh,
  Ruggenthaler, Appel, and Rubio}}]{doi:10.1021/acsphotonics.9b00768}
\bibinfo{author}{\bibfnamefont{J.}~\bibnamefont{Flick}},
  \bibinfo{author}{\bibfnamefont{D.~M.} \bibnamefont{Welakuh}},
  \bibinfo{author}{\bibfnamefont{M.}~\bibnamefont{Ruggenthaler}},
  \bibinfo{author}{\bibfnamefont{H.}~\bibnamefont{Appel}}, \bibnamefont{and}
  \bibinfo{author}{\bibfnamefont{A.}~\bibnamefont{Rubio}},
  \bibinfo{journal}{ACS Photonics} \textbf{\bibinfo{volume}{6}},
  \bibinfo{pages}{2757} (\bibinfo{year}{2019}).

\bibitem[{\citenamefont{Latini et~al.}(2019)\citenamefont{Latini, Ronca,
  De~Giovannini, Hübener, and Rubio}}]{doi:10.1021/acs.nanolett.9b00183}
\bibinfo{author}{\bibfnamefont{S.}~\bibnamefont{Latini}},
  \bibinfo{author}{\bibfnamefont{E.}~\bibnamefont{Ronca}},
  \bibinfo{author}{\bibfnamefont{U.}~\bibnamefont{De~Giovannini}},
  \bibinfo{author}{\bibfnamefont{H.}~\bibnamefont{Hübener}}, \bibnamefont{and}
  \bibinfo{author}{\bibfnamefont{A.}~\bibnamefont{Rubio}},
  \bibinfo{journal}{Nano Letters} \textbf{\bibinfo{volume}{19}},
  \bibinfo{pages}{3473} (\bibinfo{year}{2019}), \bibinfo{note}{pMID: 31046291}.

\bibitem[{\citenamefont{Flick et~al.}(2018{\natexlab{a}})\citenamefont{Flick,
  Rivera, and Narang}}]{FlickRiveraNarang}
\bibinfo{author}{\bibfnamefont{J.}~\bibnamefont{Flick}},
  \bibinfo{author}{\bibfnamefont{N.}~\bibnamefont{Rivera}}, \bibnamefont{and}
  \bibinfo{author}{\bibfnamefont{P.}~\bibnamefont{Narang}},
  \bibinfo{journal}{Nanophotonics} \textbf{\bibinfo{volume}{7}},
  \bibinfo{pages}{1479} (\bibinfo{year}{2018}{\natexlab{a}}),
  \urlprefix\url{https://doi.org/10.1515/nanoph-2018-0067}.

\bibitem[{\citenamefont{Flick and Narang}(2018)}]{PhysRevLett.121.113002}
\bibinfo{author}{\bibfnamefont{J.}~\bibnamefont{Flick}} \bibnamefont{and}
  \bibinfo{author}{\bibfnamefont{P.}~\bibnamefont{Narang}},
  \bibinfo{journal}{Phys. Rev. Lett.} \textbf{\bibinfo{volume}{121}},
  \bibinfo{pages}{113002} (\bibinfo{year}{2018}),
  \urlprefix\url{https://link.aps.org/doi/10.1103/PhysRevLett.121.113002}.

\bibitem[{\citenamefont{Garcia-Vidal et~al.}(2021)\citenamefont{Garcia-Vidal,
  Ciuti, and Ebbesen}}]{Garcia-Vidaleabd0336}
\bibinfo{author}{\bibfnamefont{F.~J.} \bibnamefont{Garcia-Vidal}},
  \bibinfo{author}{\bibfnamefont{C.}~\bibnamefont{Ciuti}}, \bibnamefont{and}
  \bibinfo{author}{\bibfnamefont{T.~W.} \bibnamefont{Ebbesen}},
  \bibinfo{journal}{Science} \textbf{\bibinfo{volume}{373}}
  (\bibinfo{year}{2021}), ISSN \bibinfo{issn}{0036-8075},
  \eprint{https://science.sciencemag.org/content/373/6551/eabd0336.full.pdf},
  \urlprefix\url{https://science.sciencemag.org/content/373/6551/eabd0336}.

\bibitem[{\citenamefont{Thomas et~al.}(2019)\citenamefont{Thomas,
  Lethuillier-Karl, Nagarajan, Vergauwe, George, Chervy, Shalabney, Devaux,
  Genet, Moran et~al.}}]{Thomas615}
\bibinfo{author}{\bibfnamefont{A.}~\bibnamefont{Thomas}},
  \bibinfo{author}{\bibfnamefont{L.}~\bibnamefont{Lethuillier-Karl}},
  \bibinfo{author}{\bibfnamefont{K.}~\bibnamefont{Nagarajan}},
  \bibinfo{author}{\bibfnamefont{R.~M.~A.} \bibnamefont{Vergauwe}},
  \bibinfo{author}{\bibfnamefont{J.}~\bibnamefont{George}},
  \bibinfo{author}{\bibfnamefont{T.}~\bibnamefont{Chervy}},
  \bibinfo{author}{\bibfnamefont{A.}~\bibnamefont{Shalabney}},
  \bibinfo{author}{\bibfnamefont{E.}~\bibnamefont{Devaux}},
  \bibinfo{author}{\bibfnamefont{C.}~\bibnamefont{Genet}},
  \bibinfo{author}{\bibfnamefont{J.}~\bibnamefont{Moran}},
  \bibnamefont{et~al.}, \bibinfo{journal}{Science}
  \textbf{\bibinfo{volume}{363}}, \bibinfo{pages}{615} (\bibinfo{year}{2019}),
  ISSN \bibinfo{issn}{0036-8075},
  \eprint{https://science.sciencemag.org/content/363/6427/615.full.pdf},
  \urlprefix\url{https://science.sciencemag.org/content/363/6427/615}.

\bibitem[{\citenamefont{Mordovina et~al.}(2020)\citenamefont{Mordovina, Bungey,
  Appel, Knowles, Rubio, and Manby}}]{PhysRevResearch.2.023262}
\bibinfo{author}{\bibfnamefont{U.}~\bibnamefont{Mordovina}},
  \bibinfo{author}{\bibfnamefont{C.}~\bibnamefont{Bungey}},
  \bibinfo{author}{\bibfnamefont{H.}~\bibnamefont{Appel}},
  \bibinfo{author}{\bibfnamefont{P.~J.} \bibnamefont{Knowles}},
  \bibinfo{author}{\bibfnamefont{A.}~\bibnamefont{Rubio}}, \bibnamefont{and}
  \bibinfo{author}{\bibfnamefont{F.~R.} \bibnamefont{Manby}},
  \bibinfo{journal}{Phys. Rev. Research} \textbf{\bibinfo{volume}{2}},
  \bibinfo{pages}{023262} (\bibinfo{year}{2020}),
  \urlprefix\url{https://link.aps.org/doi/10.1103/PhysRevResearch.2.023262}.

\bibitem[{\citenamefont{Wang et~al.}(2021{\natexlab{b}})\citenamefont{Wang,
  Neuman, Flick, and Narang}}]{doi:10.1063/5.0036283}
\bibinfo{author}{\bibfnamefont{D.~S.} \bibnamefont{Wang}},
  \bibinfo{author}{\bibfnamefont{T.}~\bibnamefont{Neuman}},
  \bibinfo{author}{\bibfnamefont{J.}~\bibnamefont{Flick}}, \bibnamefont{and}
  \bibinfo{author}{\bibfnamefont{P.}~\bibnamefont{Narang}},
  \bibinfo{journal}{The Journal of Chemical Physics}
  \textbf{\bibinfo{volume}{154}}, \bibinfo{pages}{104109}
  (\bibinfo{year}{2021}{\natexlab{b}}).

\bibitem[{\citenamefont{DePrince}(2021)}]{doi:10.1063/5.0038748}
\bibinfo{author}{\bibfnamefont{A.~E.} \bibnamefont{DePrince}},
  \bibinfo{journal}{The Journal of Chemical Physics}
  \textbf{\bibinfo{volume}{154}}, \bibinfo{pages}{094112}
  (\bibinfo{year}{2021}), \urlprefix\url{https://doi.org/10.1063/5.0038748}.

\bibitem[{\citenamefont{Haugland et~al.}(2021)\citenamefont{Haugland, Schäfer,
  Ronca, Rubio, and Koch}}]{doi:10.1063/5.0039256}
\bibinfo{author}{\bibfnamefont{T.~S.} \bibnamefont{Haugland}},
  \bibinfo{author}{\bibfnamefont{C.}~\bibnamefont{Schäfer}},
  \bibinfo{author}{\bibfnamefont{E.}~\bibnamefont{Ronca}},
  \bibinfo{author}{\bibfnamefont{A.}~\bibnamefont{Rubio}}, \bibnamefont{and}
  \bibinfo{author}{\bibfnamefont{H.}~\bibnamefont{Koch}}, \bibinfo{journal}{The
  Journal of Chemical Physics} \textbf{\bibinfo{volume}{154}},
  \bibinfo{pages}{094113} (\bibinfo{year}{2021}),
  \urlprefix\url{https://doi.org/10.1063/5.0039256}.

\bibitem[{\citenamefont{Hoffmann et~al.}(2020)\citenamefont{Hoffmann, Lacombe,
  Rubio, and Maitra}}]{doi:10.1063/5.0012723}
\bibinfo{author}{\bibfnamefont{N.~M.} \bibnamefont{Hoffmann}},
  \bibinfo{author}{\bibfnamefont{L.}~\bibnamefont{Lacombe}},
  \bibinfo{author}{\bibfnamefont{A.}~\bibnamefont{Rubio}}, \bibnamefont{and}
  \bibinfo{author}{\bibfnamefont{N.~T.} \bibnamefont{Maitra}},
  \bibinfo{journal}{The Journal of Chemical Physics}
  \textbf{\bibinfo{volume}{153}}, \bibinfo{pages}{104103}
  (\bibinfo{year}{2020}).

\bibitem[{\citenamefont{Flick and Narang}(2020)}]{doi:10.1063/5.0021033}
\bibinfo{author}{\bibfnamefont{J.}~\bibnamefont{Flick}} \bibnamefont{and}
  \bibinfo{author}{\bibfnamefont{P.}~\bibnamefont{Narang}},
  \bibinfo{journal}{The Journal of Chemical Physics}
  \textbf{\bibinfo{volume}{153}}, \bibinfo{pages}{094116}
  (\bibinfo{year}{2020}), \urlprefix\url{https://doi.org/10.1063/5.0021033}.

\bibitem[{\citenamefont{Mandal et~al.}(2020{\natexlab{b}})\citenamefont{Mandal,
  Krauss, and Huo}}]{acs.jpcb.0c03227}
\bibinfo{author}{\bibfnamefont{A.}~\bibnamefont{Mandal}},
  \bibinfo{author}{\bibfnamefont{T.~D.} \bibnamefont{Krauss}},
  \bibnamefont{and} \bibinfo{author}{\bibfnamefont{P.}~\bibnamefont{Huo}},
  \bibinfo{journal}{The Journal of Physical Chemistry B}
  \textbf{\bibinfo{volume}{124}}, \bibinfo{pages}{6321}
  (\bibinfo{year}{2020}{\natexlab{b}}), \bibinfo{note}{pMID: 32589846}.

\bibitem[{\citenamefont{Galego et~al.}(2017)\citenamefont{Galego, Garcia-Vidal,
  and Feist}}]{PhysRevLett.119.136001}
\bibinfo{author}{\bibfnamefont{J.}~\bibnamefont{Galego}},
  \bibinfo{author}{\bibfnamefont{F.~J.} \bibnamefont{Garcia-Vidal}},
  \bibnamefont{and} \bibinfo{author}{\bibfnamefont{J.}~\bibnamefont{Feist}},
  \bibinfo{journal}{Phys. Rev. Lett.} \textbf{\bibinfo{volume}{119}},
  \bibinfo{pages}{136001} (\bibinfo{year}{2017}),
  \urlprefix\url{https://link.aps.org/doi/10.1103/PhysRevLett.119.136001}.

\bibitem[{\citenamefont{Flick et~al.}(2018{\natexlab{b}})\citenamefont{Flick,
  Schäfer, Ruggenthaler, Appel, and Rubio}}]{doi:10.1021/acsphotonics.7b01279}
\bibinfo{author}{\bibfnamefont{J.}~\bibnamefont{Flick}},
  \bibinfo{author}{\bibfnamefont{C.}~\bibnamefont{Schäfer}},
  \bibinfo{author}{\bibfnamefont{M.}~\bibnamefont{Ruggenthaler}},
  \bibinfo{author}{\bibfnamefont{H.}~\bibnamefont{Appel}}, \bibnamefont{and}
  \bibinfo{author}{\bibfnamefont{A.}~\bibnamefont{Rubio}},
  \bibinfo{journal}{ACS Photonics} \textbf{\bibinfo{volume}{5}},
  \bibinfo{pages}{992} (\bibinfo{year}{2018}{\natexlab{b}}).

\bibitem[{\citenamefont{Schafer et~al.}(2018)\citenamefont{Schafer,
  Ruggenthaler, and Rubio}}]{PhysRevA.98.043801}
\bibinfo{author}{\bibfnamefont{C.}~\bibnamefont{Schafer}},
  \bibinfo{author}{\bibfnamefont{M.}~\bibnamefont{Ruggenthaler}},
  \bibnamefont{and} \bibinfo{author}{\bibfnamefont{A.}~\bibnamefont{Rubio}},
  \bibinfo{journal}{Phys. Rev. A} \textbf{\bibinfo{volume}{98}},
  \bibinfo{pages}{043801} (\bibinfo{year}{2018}),
  \urlprefix\url{https://link.aps.org/doi/10.1103/PhysRevA.98.043801}.

\bibitem[{\citenamefont{Sidler et~al.}(2020)\citenamefont{Sidler, Ruggenthaler,
  Appel, and Rubio}}]{doi:10.1021/acs.jpclett.0c01556}
\bibinfo{author}{\bibfnamefont{D.}~\bibnamefont{Sidler}},
  \bibinfo{author}{\bibfnamefont{M.}~\bibnamefont{Ruggenthaler}},
  \bibinfo{author}{\bibfnamefont{H.}~\bibnamefont{Appel}}, \bibnamefont{and}
  \bibinfo{author}{\bibfnamefont{A.}~\bibnamefont{Rubio}},
  \bibinfo{journal}{The Journal of Physical Chemistry Letters}
  \textbf{\bibinfo{volume}{11}}, \bibinfo{pages}{7525} (\bibinfo{year}{2020}),
  \bibinfo{note}{pMID: 32805122},
  \urlprefix\url{https://doi.org/10.1021/acs.jpclett.0c01556}.

\bibitem[{\citenamefont{Theophilou et~al.}(2020)\citenamefont{Theophilou, Penz,
  Ruggenthaler, and Rubio}}]{doi:10.1021/acs.jctc.0c00618}
\bibinfo{author}{\bibfnamefont{I.}~\bibnamefont{Theophilou}},
  \bibinfo{author}{\bibfnamefont{M.}~\bibnamefont{Penz}},
  \bibinfo{author}{\bibfnamefont{M.}~\bibnamefont{Ruggenthaler}},
  \bibnamefont{and} \bibinfo{author}{\bibfnamefont{A.}~\bibnamefont{Rubio}},
  \bibinfo{journal}{Journal of Chemical Theory and Computation}
  \textbf{\bibinfo{volume}{16}}, \bibinfo{pages}{6236} (\bibinfo{year}{2020}),
  \bibinfo{note}{pMID: 32816479},
  \urlprefix\url{https://doi.org/10.1021/acs.jctc.0c00618}.

\bibitem[{\citenamefont{Sidler et~al.}(2021)\citenamefont{Sidler, Schäfer,
  Ruggenthaler, and Rubio}}]{doi:10.1021/acs.jpclett.0c03436}
\bibinfo{author}{\bibfnamefont{D.}~\bibnamefont{Sidler}},
  \bibinfo{author}{\bibfnamefont{C.}~\bibnamefont{Schäfer}},
  \bibinfo{author}{\bibfnamefont{M.}~\bibnamefont{Ruggenthaler}},
  \bibnamefont{and} \bibinfo{author}{\bibfnamefont{A.}~\bibnamefont{Rubio}},
  \bibinfo{journal}{The Journal of Physical Chemistry Letters}
  \textbf{\bibinfo{volume}{12}}, \bibinfo{pages}{508} (\bibinfo{year}{2021}),
  \bibinfo{note}{pMID: 33373238},
  \urlprefix\url{https://doi.org/10.1021/acs.jpclett.0c03436}.

\bibitem[{\citenamefont{Buchholz et~al.}(2020)\citenamefont{Buchholz,
  Theophilou, Giesbertz, Ruggenthaler, and
  Rubio}}]{doi:10.1021/acs.jctc.0c00469}
\bibinfo{author}{\bibfnamefont{F.}~\bibnamefont{Buchholz}},
  \bibinfo{author}{\bibfnamefont{I.}~\bibnamefont{Theophilou}},
  \bibinfo{author}{\bibfnamefont{K.~J.~H.} \bibnamefont{Giesbertz}},
  \bibinfo{author}{\bibfnamefont{M.}~\bibnamefont{Ruggenthaler}},
  \bibnamefont{and} \bibinfo{author}{\bibfnamefont{A.}~\bibnamefont{Rubio}},
  \bibinfo{journal}{Journal of Chemical Theory and Computation}
  \textbf{\bibinfo{volume}{16}}, \bibinfo{pages}{5601} (\bibinfo{year}{2020}),
  \bibinfo{note}{pMID: 32692551},
  \urlprefix\url{https://doi.org/10.1021/acs.jctc.0c00469}.

\bibitem[{\citenamefont{Tokatly}(2018)}]{PhysRevB.98.235123}
\bibinfo{author}{\bibfnamefont{I.~V.} \bibnamefont{Tokatly}},
  \bibinfo{journal}{Phys. Rev. B} \textbf{\bibinfo{volume}{98}},
  \bibinfo{pages}{235123} (\bibinfo{year}{2018}),
  \urlprefix\url{https://link.aps.org/doi/10.1103/PhysRevB.98.235123}.

\bibitem[{\citenamefont{Rivera et~al.}(2019)\citenamefont{Rivera, Flick, and
  Narang}}]{PhysRevLett.122.193603}
\bibinfo{author}{\bibfnamefont{N.}~\bibnamefont{Rivera}},
  \bibinfo{author}{\bibfnamefont{J.}~\bibnamefont{Flick}}, \bibnamefont{and}
  \bibinfo{author}{\bibfnamefont{P.}~\bibnamefont{Narang}},
  \bibinfo{journal}{Phys. Rev. Lett.} \textbf{\bibinfo{volume}{122}},
  \bibinfo{pages}{193603} (\bibinfo{year}{2019}),
  \urlprefix\url{https://link.aps.org/doi/10.1103/PhysRevLett.122.193603}.

\bibitem[{\citenamefont{Szidarovszky et~al.}(2020)\citenamefont{Szidarovszky,
  Hal{\'{a}}sz, and Vib{\'{o}}k}}]{Szidarovszky_2020}
\bibinfo{author}{\bibfnamefont{T.}~\bibnamefont{Szidarovszky}},
  \bibinfo{author}{\bibfnamefont{G.~J.} \bibnamefont{Hal{\'{a}}sz}},
  \bibnamefont{and}
  \bibinfo{author}{\bibfnamefont{{\'{A}}.}~\bibnamefont{Vib{\'{o}}k}},
  \bibinfo{journal}{New Journal of Physics} \textbf{\bibinfo{volume}{22}},
  \bibinfo{pages}{053001} (\bibinfo{year}{2020}),
  \urlprefix\url{https://doi.org/10.1088/1367-2630/ab8264}.

\bibitem[{\citenamefont{Cederbaum}(2021)}]{doi:10.1021/acs.jpclett.1c01570}
\bibinfo{author}{\bibfnamefont{L.~S.} \bibnamefont{Cederbaum}},
  \bibinfo{journal}{The Journal of Physical Chemistry Letters}
  \textbf{\bibinfo{volume}{12}}, \bibinfo{pages}{6056} (\bibinfo{year}{2021}),
  \bibinfo{note}{pMID: 34165990},
  \urlprefix\url{https://doi.org/10.1021/acs.jpclett.1c01570}.

\bibitem[{\citenamefont{Thomas et~al.}(2016)\citenamefont{Thomas, George,
  Shalabney, Dryzhakov, Varma, Moran, Chervy, Zhong, Devaux, Genet
  et~al.}}]{https://doi.org/10.1002/anie.201605504}
\bibinfo{author}{\bibfnamefont{A.}~\bibnamefont{Thomas}},
  \bibinfo{author}{\bibfnamefont{J.}~\bibnamefont{George}},
  \bibinfo{author}{\bibfnamefont{A.}~\bibnamefont{Shalabney}},
  \bibinfo{author}{\bibfnamefont{M.}~\bibnamefont{Dryzhakov}},
  \bibinfo{author}{\bibfnamefont{S.~J.} \bibnamefont{Varma}},
  \bibinfo{author}{\bibfnamefont{J.}~\bibnamefont{Moran}},
  \bibinfo{author}{\bibfnamefont{T.}~\bibnamefont{Chervy}},
  \bibinfo{author}{\bibfnamefont{X.}~\bibnamefont{Zhong}},
  \bibinfo{author}{\bibfnamefont{E.}~\bibnamefont{Devaux}},
  \bibinfo{author}{\bibfnamefont{C.}~\bibnamefont{Genet}},
  \bibnamefont{et~al.}, \bibinfo{journal}{Angewandte Chemie International
  Edition} \textbf{\bibinfo{volume}{55}}, \bibinfo{pages}{11462}
  (\bibinfo{year}{2016}),
  \eprint{https://onlinelibrary.wiley.com/doi/pdf/10.1002/anie.201605504},
  \urlprefix\url{https://onlinelibrary.wiley.com/doi/abs/10.1002/anie.201605504}.

\bibitem[{\citenamefont{Puchalski et~al.}(2019)\citenamefont{Puchalski, Komasa,
  Czachorowski, and Pachucki}}]{PhysRevLett.122.103003}
\bibinfo{author}{\bibfnamefont{M.}~\bibnamefont{Puchalski}},
  \bibinfo{author}{\bibfnamefont{J.}~\bibnamefont{Komasa}},
  \bibinfo{author}{\bibfnamefont{P.}~\bibnamefont{Czachorowski}},
  \bibnamefont{and} \bibinfo{author}{\bibfnamefont{K.}~\bibnamefont{Pachucki}},
  \bibinfo{journal}{Phys. Rev. Lett.} \textbf{\bibinfo{volume}{122}},
  \bibinfo{pages}{103003} (\bibinfo{year}{2019}),
  \urlprefix\url{https://link.aps.org/doi/10.1103/PhysRevLett.122.103003}.

\bibitem[{\citenamefont{H\"olsch et~al.}(2019)\citenamefont{H\"olsch, Beyer,
  Salumbides, Eikema, Ubachs, Jungen, and Merkt}}]{PhysRevLett.122.103002}
\bibinfo{author}{\bibfnamefont{N.}~\bibnamefont{H\"olsch}},
  \bibinfo{author}{\bibfnamefont{M.}~\bibnamefont{Beyer}},
  \bibinfo{author}{\bibfnamefont{E.~J.} \bibnamefont{Salumbides}},
  \bibinfo{author}{\bibfnamefont{K.~S.~E.} \bibnamefont{Eikema}},
  \bibinfo{author}{\bibfnamefont{W.}~\bibnamefont{Ubachs}},
  \bibinfo{author}{\bibfnamefont{C.}~\bibnamefont{Jungen}}, \bibnamefont{and}
  \bibinfo{author}{\bibfnamefont{F.}~\bibnamefont{Merkt}},
  \bibinfo{journal}{Phys. Rev. Lett.} \textbf{\bibinfo{volume}{122}},
  \bibinfo{pages}{103002} (\bibinfo{year}{2019}),
  \urlprefix\url{https://link.aps.org/doi/10.1103/PhysRevLett.122.103002}.

\bibitem[{\citenamefont{Kohn and Sham}(1965)}]{PhysRev.140.A1133}
\bibinfo{author}{\bibfnamefont{W.}~\bibnamefont{Kohn}} \bibnamefont{and}
  \bibinfo{author}{\bibfnamefont{L.~J.} \bibnamefont{Sham}},
  \bibinfo{journal}{Phys. Rev.} \textbf{\bibinfo{volume}{140}},
  \bibinfo{pages}{A1133} (\bibinfo{year}{1965}),
  \urlprefix\url{https://link.aps.org/doi/10.1103/PhysRev.140.A1133}.

\bibitem[{\citenamefont{Mitroy et~al.}(2013)\citenamefont{Mitroy, Bubin,
  Horiuchi, Suzuki, Adamowicz, Cencek, Szalewicz, Komasa, Blume, and
  Varga}}]{RevModPhys.85.693}
\bibinfo{author}{\bibfnamefont{J.}~\bibnamefont{Mitroy}},
  \bibinfo{author}{\bibfnamefont{S.}~\bibnamefont{Bubin}},
  \bibinfo{author}{\bibfnamefont{W.}~\bibnamefont{Horiuchi}},
  \bibinfo{author}{\bibfnamefont{Y.}~\bibnamefont{Suzuki}},
  \bibinfo{author}{\bibfnamefont{L.}~\bibnamefont{Adamowicz}},
  \bibinfo{author}{\bibfnamefont{W.}~\bibnamefont{Cencek}},
  \bibinfo{author}{\bibfnamefont{K.}~\bibnamefont{Szalewicz}},
  \bibinfo{author}{\bibfnamefont{J.}~\bibnamefont{Komasa}},
  \bibinfo{author}{\bibfnamefont{D.}~\bibnamefont{Blume}}, \bibnamefont{and}
  \bibinfo{author}{\bibfnamefont{K.}~\bibnamefont{Varga}},
  \bibinfo{journal}{Rev. Mod. Phys.} \textbf{\bibinfo{volume}{85}},
  \bibinfo{pages}{693} (\bibinfo{year}{2013}),
  \urlprefix\url{https://link.aps.org/doi/10.1103/RevModPhys.85.693}.

\bibitem[{\citenamefont{Suzuki et~al.}(1998)\citenamefont{Suzuki, Suzuki, and
  Varga}}]{suzuki1998stochastic}
\bibinfo{author}{\bibfnamefont{Y.}~\bibnamefont{Suzuki}},
  \bibinfo{author}{\bibfnamefont{M.}~\bibnamefont{Suzuki}}, \bibnamefont{and}
  \bibinfo{author}{\bibfnamefont{K.}~\bibnamefont{Varga}},
  \emph{\bibinfo{title}{Stochastic variational approach to quantum-mechanical
  few-body problems}}, vol.~\bibinfo{volume}{54} (\bibinfo{publisher}{Springer
  Science \& Business Media}, \bibinfo{year}{1998}).

\bibitem[{\citenamefont{Cioslowski and
  Strasburger}(2017)}]{doi:10.1063/1.4974273}
\bibinfo{author}{\bibfnamefont{J.}~\bibnamefont{Cioslowski}} \bibnamefont{and}
  \bibinfo{author}{\bibfnamefont{K.}~\bibnamefont{Strasburger}},
  \bibinfo{journal}{The Journal of Chemical Physics}
  \textbf{\bibinfo{volume}{146}}, \bibinfo{pages}{044308}
  (\bibinfo{year}{2017}), \urlprefix\url{https://doi.org/10.1063/1.4974273}.

\bibitem[{\citenamefont{Zaklama et~al.}(2019)\citenamefont{Zaklama, Zhang,
  Rowan, Schatzki, Suzuki, and Varga}}]{Zaklama2019}
\bibinfo{author}{\bibfnamefont{T.}~\bibnamefont{Zaklama}},
  \bibinfo{author}{\bibfnamefont{D.}~\bibnamefont{Zhang}},
  \bibinfo{author}{\bibfnamefont{K.}~\bibnamefont{Rowan}},
  \bibinfo{author}{\bibfnamefont{L.}~\bibnamefont{Schatzki}},
  \bibinfo{author}{\bibfnamefont{Y.}~\bibnamefont{Suzuki}}, \bibnamefont{and}
  \bibinfo{author}{\bibfnamefont{K.}~\bibnamefont{Varga}},
  \bibinfo{journal}{Few-Body Systems} \textbf{\bibinfo{volume}{61}},
  \bibinfo{pages}{6} (\bibinfo{year}{2019}), ISSN \bibinfo{issn}{1432-5411},
  \urlprefix\url{https://doi.org/10.1007/s00601-019-1539-3}.

\bibitem[{\citenamefont{Zhang et~al.}(2015)\citenamefont{Zhang, Kidd, and
  Varga}}]{Zhang2015}
\bibinfo{author}{\bibfnamefont{D.~K.} \bibnamefont{Zhang}},
  \bibinfo{author}{\bibfnamefont{D.~W.} \bibnamefont{Kidd}}, \bibnamefont{and}
  \bibinfo{author}{\bibfnamefont{K.}~\bibnamefont{Varga}},
  \bibinfo{journal}{Nano Letters} \textbf{\bibinfo{volume}{15}},
  \bibinfo{pages}{7002} (\bibinfo{year}{2015}), ISSN \bibinfo{issn}{1530-6984},
  \urlprefix\url{https://doi.org/10.1021/acs.nanolett.5b03009}.

\bibitem[{\citenamefont{Kidd et~al.}(2016)\citenamefont{Kidd, Zhang, and
  Varga}}]{PhysRevB.93.125423}
\bibinfo{author}{\bibfnamefont{D.~W.} \bibnamefont{Kidd}},
  \bibinfo{author}{\bibfnamefont{D.~K.} \bibnamefont{Zhang}}, \bibnamefont{and}
  \bibinfo{author}{\bibfnamefont{K.}~\bibnamefont{Varga}},
  \bibinfo{journal}{Phys. Rev. B} \textbf{\bibinfo{volume}{93}},
  \bibinfo{pages}{125423} (\bibinfo{year}{2016}),
  \urlprefix\url{https://link.aps.org/doi/10.1103/PhysRevB.93.125423}.

\bibitem[{\citenamefont{Riva et~al.}(2000)\citenamefont{Riva, Peeters, and
  Varga}}]{PhysRevB.61.13873}
\bibinfo{author}{\bibfnamefont{C.}~\bibnamefont{Riva}},
  \bibinfo{author}{\bibfnamefont{F.~M.} \bibnamefont{Peeters}},
  \bibnamefont{and} \bibinfo{author}{\bibfnamefont{K.}~\bibnamefont{Varga}},
  \bibinfo{journal}{Phys. Rev. B} \textbf{\bibinfo{volume}{61}},
  \bibinfo{pages}{13873} (\bibinfo{year}{2000}),
  \urlprefix\url{https://link.aps.org/doi/10.1103/PhysRevB.61.13873}.

\bibitem[{\citenamefont{Varga}(1999)}]{PhysRevLett.83.5471}
\bibinfo{author}{\bibfnamefont{K.}~\bibnamefont{Varga}},
  \bibinfo{journal}{Phys. Rev. Lett.} \textbf{\bibinfo{volume}{83}},
  \bibinfo{pages}{5471} (\bibinfo{year}{1999}),
  \urlprefix\url{https://link.aps.org/doi/10.1103/PhysRevLett.83.5471}.

\bibitem[{\citenamefont{Usukura et~al.}(1999)\citenamefont{Usukura, Suzuki, and
  Varga}}]{PhysRevB.59.5652}
\bibinfo{author}{\bibfnamefont{J.}~\bibnamefont{Usukura}},
  \bibinfo{author}{\bibfnamefont{Y.}~\bibnamefont{Suzuki}}, \bibnamefont{and}
  \bibinfo{author}{\bibfnamefont{K.}~\bibnamefont{Varga}},
  \bibinfo{journal}{Phys. Rev. B} \textbf{\bibinfo{volume}{59}},
  \bibinfo{pages}{5652} (\bibinfo{year}{1999}),
  \urlprefix\url{https://link.aps.org/doi/10.1103/PhysRevB.59.5652}.

\bibitem[{\citenamefont{Varga et~al.}(2001)\citenamefont{Varga, Navratil,
  Usukura, and Suzuki}}]{PhysRevB.63.205308}
\bibinfo{author}{\bibfnamefont{K.}~\bibnamefont{Varga}},
  \bibinfo{author}{\bibfnamefont{P.}~\bibnamefont{Navratil}},
  \bibinfo{author}{\bibfnamefont{J.}~\bibnamefont{Usukura}}, \bibnamefont{and}
  \bibinfo{author}{\bibfnamefont{Y.}~\bibnamefont{Suzuki}},
  \bibinfo{journal}{Phys. Rev. B} \textbf{\bibinfo{volume}{63}},
  \bibinfo{pages}{205308} (\bibinfo{year}{2001}),
  \urlprefix\url{https://link.aps.org/doi/10.1103/PhysRevB.63.205308}.

\bibitem[{\citenamefont{Rokaj et~al.}(2018)\citenamefont{Rokaj, Welakuh,
  Ruggenthaler, and Rubio}}]{Rokaj_2018}
\bibinfo{author}{\bibfnamefont{V.}~\bibnamefont{Rokaj}},
  \bibinfo{author}{\bibfnamefont{D.~M.} \bibnamefont{Welakuh}},
  \bibinfo{author}{\bibfnamefont{M.}~\bibnamefont{Ruggenthaler}},
  \bibnamefont{and} \bibinfo{author}{\bibfnamefont{A.}~\bibnamefont{Rubio}},
  \bibinfo{journal}{Journal of Physics B: Atomic, Molecular and Optical
  Physics} \textbf{\bibinfo{volume}{51}}, \bibinfo{pages}{034005}
  (\bibinfo{year}{2018}),
  \urlprefix\url{https://doi.org/10.1088/1361-6455/aa9c99}.

\bibitem[{\citenamefont{Ruggenthaler et~al.}(2014)\citenamefont{Ruggenthaler,
  Flick, Pellegrini, Appel, Tokatly, and Rubio}}]{PhysRevA.90.012508}
\bibinfo{author}{\bibfnamefont{M.}~\bibnamefont{Ruggenthaler}},
  \bibinfo{author}{\bibfnamefont{J.}~\bibnamefont{Flick}},
  \bibinfo{author}{\bibfnamefont{C.}~\bibnamefont{Pellegrini}},
  \bibinfo{author}{\bibfnamefont{H.}~\bibnamefont{Appel}},
  \bibinfo{author}{\bibfnamefont{I.~V.} \bibnamefont{Tokatly}},
  \bibnamefont{and} \bibinfo{author}{\bibfnamefont{A.}~\bibnamefont{Rubio}},
  \bibinfo{journal}{Phys. Rev. A} \textbf{\bibinfo{volume}{90}},
  \bibinfo{pages}{012508} (\bibinfo{year}{2014}),
  \urlprefix\url{https://link.aps.org/doi/10.1103/PhysRevA.90.012508}.

\bibitem[{Sup()}]{Supp}
 (????), \bibinfo{note}{see Supplemental Material}.

\bibitem[{\citenamefont{Kukulin and {Krasnopol'sky}}(1977)}]{kukulin77}
\bibinfo{author}{\bibfnamefont{V.~I.} \bibnamefont{Kukulin}} \bibnamefont{and}
  \bibinfo{author}{\bibfnamefont{V.~M.} \bibnamefont{{Krasnopol'sky}}},
  \bibinfo{journal}{J.~Phys.~G} \textbf{\bibinfo{volume}{3}},
  \bibinfo{pages}{795} (\bibinfo{year}{1977}).

\bibitem[{\citenamefont{Huang et~al.}(2021)\citenamefont{Huang, Ahrens, Beutel,
  and Varga}}]{henry}
\bibinfo{author}{\bibfnamefont{C.}~\bibnamefont{Huang}},
  \bibinfo{author}{\bibfnamefont{A.}~\bibnamefont{Ahrens}},
  \bibinfo{author}{\bibfnamefont{M.}~\bibnamefont{Beutel}}, \bibnamefont{and}
  \bibinfo{author}{\bibfnamefont{K.}~\bibnamefont{Varga}},
  \emph{\bibinfo{title}{Two electrons in harmonic confinement coupled to light
  in a cavity}} (\bibinfo{year}{2021}), \eprint{2108.01702}.

\bibitem[{\citenamefont{Emmanuele et~al.}(2020)\citenamefont{Emmanuele, Sich,
  Kyriienko, Shahnazaryan, Withers, Catanzaro, Walker, Benimetskiy, Skolnick,
  Tartakovskii et~al.}}]{Emmanuele2020}
\bibinfo{author}{\bibfnamefont{R.~P.~A.} \bibnamefont{Emmanuele}},
  \bibinfo{author}{\bibfnamefont{M.}~\bibnamefont{Sich}},
  \bibinfo{author}{\bibfnamefont{O.}~\bibnamefont{Kyriienko}},
  \bibinfo{author}{\bibfnamefont{V.}~\bibnamefont{Shahnazaryan}},
  \bibinfo{author}{\bibfnamefont{F.}~\bibnamefont{Withers}},
  \bibinfo{author}{\bibfnamefont{A.}~\bibnamefont{Catanzaro}},
  \bibinfo{author}{\bibfnamefont{P.~M.} \bibnamefont{Walker}},
  \bibinfo{author}{\bibfnamefont{F.~A.} \bibnamefont{Benimetskiy}},
  \bibinfo{author}{\bibfnamefont{M.~S.} \bibnamefont{Skolnick}},
  \bibinfo{author}{\bibfnamefont{A.~I.} \bibnamefont{Tartakovskii}},
  \bibnamefont{et~al.}, \bibinfo{journal}{Nature Communications}
  \textbf{\bibinfo{volume}{11}}, \bibinfo{pages}{3589} (\bibinfo{year}{2020}),
  ISSN \bibinfo{issn}{2041-1723},
  \urlprefix\url{https://doi.org/10.1038/s41467-020-17340-z}.

\bibitem[{\citenamefont{Dhara et~al.}(2018)\citenamefont{Dhara, Chakraborty,
  Goodfellow, Qiu, O'Loughlin, Wicks, Bhattacharjee, and
  Vamivakas}}]{Dhara2018}
\bibinfo{author}{\bibfnamefont{S.}~\bibnamefont{Dhara}},
  \bibinfo{author}{\bibfnamefont{C.}~\bibnamefont{Chakraborty}},
  \bibinfo{author}{\bibfnamefont{K.~M.} \bibnamefont{Goodfellow}},
  \bibinfo{author}{\bibfnamefont{L.}~\bibnamefont{Qiu}},
  \bibinfo{author}{\bibfnamefont{T.~A.} \bibnamefont{O'Loughlin}},
  \bibinfo{author}{\bibfnamefont{G.~W.} \bibnamefont{Wicks}},
  \bibinfo{author}{\bibfnamefont{S.}~\bibnamefont{Bhattacharjee}},
  \bibnamefont{and} \bibinfo{author}{\bibfnamefont{A.~N.}
  \bibnamefont{Vamivakas}}, \bibinfo{journal}{Nature Physics}
  \textbf{\bibinfo{volume}{14}}, \bibinfo{pages}{130} (\bibinfo{year}{2018}),
  ISSN \bibinfo{issn}{1745-2481},
  \urlprefix\url{https://doi.org/10.1038/nphys4303}.

\bibitem[{\citenamefont{Dufferwiel et~al.}(2017)\citenamefont{Dufferwiel,
  Lyons, Solnyshkov, Trichet, Withers, Schwarz, Malpuech, Smith, Novoselov,
  Skolnick et~al.}}]{Dufferwiel2017}
\bibinfo{author}{\bibfnamefont{S.}~\bibnamefont{Dufferwiel}},
  \bibinfo{author}{\bibfnamefont{T.~P.} \bibnamefont{Lyons}},
  \bibinfo{author}{\bibfnamefont{D.~D.} \bibnamefont{Solnyshkov}},
  \bibinfo{author}{\bibfnamefont{A.~A.~P.} \bibnamefont{Trichet}},
  \bibinfo{author}{\bibfnamefont{F.}~\bibnamefont{Withers}},
  \bibinfo{author}{\bibfnamefont{S.}~\bibnamefont{Schwarz}},
  \bibinfo{author}{\bibfnamefont{G.}~\bibnamefont{Malpuech}},
  \bibinfo{author}{\bibfnamefont{J.~M.} \bibnamefont{Smith}},
  \bibinfo{author}{\bibfnamefont{K.~S.} \bibnamefont{Novoselov}},
  \bibinfo{author}{\bibfnamefont{M.~S.} \bibnamefont{Skolnick}},
  \bibnamefont{et~al.}, \bibinfo{journal}{Nature Photonics}
  \textbf{\bibinfo{volume}{11}}, \bibinfo{pages}{497} (\bibinfo{year}{2017}),
  ISSN \bibinfo{issn}{1749-4893},
  \urlprefix\url{https://doi.org/10.1038/nphoton.2017.125}.

\bibitem[{\citenamefont{Sidler et~al.}(2017)\citenamefont{Sidler, Back, Cotlet,
  Srivastava, Fink, Kroner, Demler, and Imamoglu}}]{Sidler2017}
\bibinfo{author}{\bibfnamefont{M.}~\bibnamefont{Sidler}},
  \bibinfo{author}{\bibfnamefont{P.}~\bibnamefont{Back}},
  \bibinfo{author}{\bibfnamefont{O.}~\bibnamefont{Cotlet}},
  \bibinfo{author}{\bibfnamefont{A.}~\bibnamefont{Srivastava}},
  \bibinfo{author}{\bibfnamefont{T.}~\bibnamefont{Fink}},
  \bibinfo{author}{\bibfnamefont{M.}~\bibnamefont{Kroner}},
  \bibinfo{author}{\bibfnamefont{E.}~\bibnamefont{Demler}}, \bibnamefont{and}
  \bibinfo{author}{\bibfnamefont{A.}~\bibnamefont{Imamoglu}},
  \bibinfo{journal}{Nature Physics} \textbf{\bibinfo{volume}{13}},
  \bibinfo{pages}{255} (\bibinfo{year}{2017}), ISSN \bibinfo{issn}{1745-2481},
  \urlprefix\url{https://doi.org/10.1038/nphys3949}.

\bibitem[{\citenamefont{Jaynes and Cummings}(1963)}]{1443594}
\bibinfo{author}{\bibfnamefont{E.}~\bibnamefont{Jaynes}} \bibnamefont{and}
  \bibinfo{author}{\bibfnamefont{F.}~\bibnamefont{Cummings}},
  \bibinfo{journal}{Proceedings of the IEEE} \textbf{\bibinfo{volume}{51}},
  \bibinfo{pages}{89} (\bibinfo{year}{1963}).

\end{thebibliography}

\end{document}